\documentclass[twocolumn,twocolappendix,tighten]{aastex631}
\usepackage{hyperref}
\usepackage{times, amsmath, amssymb, amsfonts, latexsym}
\usepackage{fleqn, multirow, graphicx, color, enumitem}
\usepackage{array, mathtools,subfigure, esint,boondox-cal}
\usepackage{soul}
\newcommand{\uat}[2]{\href{http://astrothesaurus.org/uat/#2}{#1 (#2)}}

\makeatletter
\@for\next:={int,iint,iiint,iiiint,dotsint,oint,oiint,sqint,sqiint,
	ointctrclockwise,ointclockwise,varointclockwise,varointctrclockwise,
	fint,varoiint,landupint,landdownint}\do{%
	\expandafter\edef\csname\next\endcsname{%
		\noexpand\DOTSI
		\expandafter\noexpand\csname\next op\endcsname
		\noexpand\ilimits@
	}%
}
\makeatother
\shorttitle{Scale and environment dependence of matter clustering}
\shortauthors{Wang \& He}
\begin{document}
\title{Simultaneous Dependence of Matter Clustering on Scale and Environment}

\author[0000-0003-4064-417X]{Yun Wang}
\affiliation{College of Physics, Jilin University, Changchun 130012, P.R. China.}

\author[0000-0001-7767-6154]{Ping He}
\affiliation{College of Physics, Jilin University, Changchun 130012, P.R. China.}
\affiliation{Center for High Energy Physics, Peking University, Beijing 100871, P.R. China.}

\correspondingauthor{Ping He}
\email{hep@jlu.edu.cn}
\begin{abstract}
\noindent In this work, we propose new statistical tools that are capable of characterizing the simultaneous dependence of dark matter and gas clustering on the scale and the density environment, and these are the environment-dependent wavelet power spectrum (env-WPS), the environment-dependent bias function (env-bias), and the environment-dependent wavelet cross-correlation function (env-WCC). These statistics are applied to the dark matter and baryonic gas density fields of the \texttt{TNG100-1} simulation at redshifts of $z=3.0$-$0.0$, and to  \texttt{Illustris-1} and \texttt{SIMBA} at $z=0$. The measurements of the env-WPSs suggest that the clustering strengths of both the dark matter and the gas increase with increasing density, while that of a Gaussian field shows no density dependence. By measuring the env-bias and env-WCC, we find that they vary significantly with the environment, scale, and redshift. A noteworthy feature is that at $z=0.0$, the gas is less biased in denser environments of $\Delta \gtrsim 10$ around $3 \ h\mathrm{Mpc}^{-1}$, due to the gas reaccretion caused by the decreased AGN feedback strength at lower redshifts. We also find that the gas correlates more tightly with the dark matter in both the most dense and underdense environments than in other environments at all epochs. Even at $z=0$, the env-WCC is greater than $0.9$ in $\Delta \gtrsim 200$ and $\Delta \lesssim 0.1$ at scales of $k \lesssim 10 \ h\mathrm{Mpc}^{-1}$. In summary, our results support the local density environment having a non-negligible impact on the deviations between dark matter and gas distributions up to large scales.
\end{abstract}

\section{Introduction}
\label{sec:intro}

In the standard paradigm of galaxy formation, small density fluctuations of baryonic gas start to grow and develop together with those of dark matter under the gravitational instability following recombination. At later times, dark matter fluctuations, once they exceed some threshold, will collapse into virialized halos, within which partial gas condenses and cools to form galaxies \citep{Mo2010}. Therefore, galaxies are biased discrete tracers of the underlying matter distribution, while the gas, as the continuously fluctuating medium, may better trace the matter distribution. On the simulation side, many studies have confirmed that the baryonic gas follows the underlying dark matter distribution on large scales quite well. For example, the power spectrum of the gas density field is very close to that of dark matter \citep[e.g.,][]{Cui2017, Springel2018}, the spatial distributions of the dark matter and baryonic gas are highly correlated with each other over a wide range of scales \citep[e.g.,][]{Yang2020, Yang2021}, and the gas alone can be used to classify the cosmic web in an un-biased manner \citep[e.g.,][]{Cui2018}. Additionally, the intergalactic gas distribution is receiving increasing attention in observations. Measuring the large-scale structures of the universe traced by the gas is the goal of the upcoming surveys, such as the Canadian Hydrogen Intensity Mapping Experiment \citep[CHIME;][]{Bandura2014}, the Hydrogen Intensity and Real-time Analysis eXperiment \citep[HIRAX;][]{Newburgh2016}, and the Square Kilometre Array \citep[SKA;][]{Bacon2020}.

As the scale becomes smaller, galaxy formation processes, such as gas cooling, star formation, and feedbacks, have an increasing influence on the matter clustering, thereby resulting in a significant bias between the gas and the dark matter. In particular, the clustering of the gas is strongly suppressed up to scales a few times $0.1 \ h\mathrm{Mpc}^{-1}$, mainly due to the active galactic nuclei (AGN) feedback, which heats and ejects gas from halos \citep{vanDaalen2019}, while the star formation consumes the available gas on galactic scales. Due to the combined effects of nonlinear gravitational evolution and complex baryonic physical processes, the matter distribution becomes highly non-Gaussian at small scales and low redshifts. Moreover, it is well known that there is a strong relationship between a galaxy's properties and its density environments. For instance, massive and quenched galaxies tend to occur in high-density environments \citep[e.g.][]{Hoyle2012, Moorman2016}, and the star formation rates of galaxies increase with environmental density at redshifts $z\gtrsim 1$, but decrease with environmental density at redshifts $z\lesssim 1$ \citep[e.g.][]{Cooper2008, Wang2018, Hwang2019}. The environmental dependence of galactic properties may imply two aspects. First, galaxies in high-density environments interact more frequently and therefore strip their gas faster than those in low-density environments. Second, feedback (e.g. AGN feedback) efficiency varies in different density environments. For example, \citet{Miraghaei2020} found in Sloan Digital Sky Survey Data Release 7 that underdense regions have a higher fraction of thermal-mode AGNs than overdense regions, for the massive red galaxies, whereas kinetic-mode AGNs prefer to reside in denser regions.

According to the above statements, deviations of the spatial clustering between the gas and the dark matter are expected to depend on both the scale and the density environment. Quantifying such simultaneous dependence on the scale and the environment will hopefully deepen our understanding of the extent to which baryonic gas follows the dark matter. To achieve this goal, statistical tools that can give the frequency information, while retaining the spatial information, are required. However, the traditional two-point correlation function or its Fourier counterpart, the power spectrum, is only a univariate function of scale and totally insensitive to non-Gaussianity, so it is not up to the task. A variety of advanced statistics beyond the power spectrum have been previously developed, including but not limited to the bispectrum, which is a third-order spectrum \citep[see e.g.][]{Sefusatti2006, Foreman2020}, the position-dependent power spectrum, based on the windowed Fourier transform \citep{Chiang2014}, the sliced correlation function \citep{Neyrinck2018}, and the k-nearest neighbor cumulative distribution functions \citep{Banerjee2020}. Since the processes associated with galaxy formation are very complex and poorly understood, predictions from different statistics are vital to improving our knowledge of galaxy formation.

As a bivariate function of the scale and space, wavelet transforms are particularly suitable for the issue we are interested in. With a basis function (i.e. a ``wavelet") that is well localized, both in the real domain and the frequency domain, the wavelet transform acts like a ``mathematical microscope", which allows us to zoom in on fine structures of the universe at various scales and locations \citep[e.g.][]{Kaiser1994, Addison2017}. There are two major types of wavelet transforms---the discrete wavelet transform (DWT) and the continuous wavelet transform (CWT)---both of which are widely employed in cosmology \citep[see e.g.][]{Martinez1993, Pando1996, Fang2000, Starck2004, Liu2008, Zhang2011, Arnalte-Mur2012, daCunha2018, Shi2018, Copi2019, Wang2021b}. Besides, there are also some other wavelet transform variants, e.g. the wavelet scattering transform \citep{Mallat2012}, which performs two operations repeatedly---(1) wavelet convolution and (2) modulus---and has been proved to be superior to the Fourier power spectrum for cosmological parameter inference \citep[e.g.][]{Allys2020, Cheng2020, Valogiannis2022}. However, using wavelets to investigate the scale and environmental dependence of the matter clustering seems to be neglected in most previous studies. Due to the arbitrary scale choices and translational invariance of the CWT \citep{Addison2017}, we will employ it to analyze the deviation between the dark matter and gas distributions, and quantify its dependence on both the scale and density environment.

The most criticized aspect of the CWT is that its computation consumes more time and resources than that of the DWT. However, the Fourier convolution theorem allows us to compute the CWT quickly, by utilizing a fast Fourier transform (FFT) with a complexity of $O(N\log N)$ at each scale \citep{Torrence1998}. In addition to this, much research has been dedicated to the fast implementation of the CWT in the real domain \citep{Rioul1991, Unser1994, Berkner1997, Vrhel1997, Munoz2002, Omachi2007, Patil2009, Aeizumi2019}. For example, by decomposing the wavelet function and signal into B-spline bases, \citet{Munoz2002} converted the CWT into a convolution of two B-splines, which could be expressed analytically and had a lower number of operations. \citet{Omachi2007} achieved a fast computation of the CWT by representing the wavelet as a polynomial within its compact support interval. The most recent algorithm, proposed by \citet{Aeizumi2019}, approximates the mother wavelet as piecewise polynomials. Although it is claimed that all these algorithms are better than FFT, they are restricted to 1D and special wavelets. Therefore, the efficiency of these methods might not be guaranteed for the general wavelet functions or for high dimensions. In the present work, the fast computation of CWT is implemented by the FFT.

In a previous work \citep{Wang2021a}, we developed a new method for designing continuous wavelets, in which those wavelets are constructed by taking the first derivative of the smoothing function, e.g. the Gaussian function, with respect to the positively defined scale parameter. The convenience of this method is due to the original signal being reconstructed through a single integral of the continuous wavelet coefficients. Furthermore, \citet{Wang2021b} introduced wavelet-based statistical quantities, including the wavelet power spectrum (WPS), wavelet cross-correlation (WCC), and wavelet bicoherence. By evenly splitting the whole simulation box into subcubes, \citet{Wang2021b} suggest that the properties of the matter clustering depend on the local mean density of the subcube. However, we realize that this is not a wise way to divide the environment and that it does not fully reflect the power of the wavelet techniques. On the one hand, the cubic environment could not isolate a special structure, say a halo, from its surroundings. On the other hand, the CWT varies with the spatial position at a fixed scale, and therefore with the local density at that location. Following this idea, we can build more flexible wavelet statistics that depend directly on the local density and scale. In this study, we propose the environment-dependent wavelet power spectrum (env-WPS) and the environment-dependent wavelet cross-correlation function (env-WCC). The env-WPS measures the strength of the matter clustering across different scales within a density interval, while the env-WCC measures the statistical coherence of the spatial distributions between the gas and the dark matter. The environment-dependent bias function (env-bias) reflects the scale- and environment-dependent bias between the gas and the dark matter. 

The environmental dependence requires wavelets with good spatial resolution, and the scale dependence requires wavelets with good frequency resolution. As a result, a wavelet that can achieve a good trade-off between these two resolutions is necessary. To achieve this, the analytic wavelet function that we used here is derived from a Gaussian function weighted by a cosine function, which has a better frequency resolution than the Gaussian-derived wavelet (GDW) used in previous works \citep{Wang2021a, Wang2021b}, and also maintains a reasonable spatial resolution. In this study, we apply the env-WPS and the env-WCC to the density fields of three cosmological simulations: Illustris \citep{Nelson2015}, IllustrisTNG \citep{Nelson2019} and SIMBA \citep{Dave2019}.

The rest of this work is structured as follows. We briefly describe the simulations we used in Section 2, and introduce the fundamental theories in Section 3. Our main results regarding the dependence of the matter clustering on both the scale and the environment are given in Section 4. Finally, in Section 5, we discuss and summarize our main findings and present the conclusions.

For convenience of the readers, in Table \ref{tab:notations}, we list the wavelet-related acronyms used in our paper, with their meanings.
\begin{table}
	\centering
	\caption{The Wavelet-related Acronyms Used in the Paper, with Their Meanings Explained}
	\label{tab:notations}
	\begin{tabular}{ll}
		\toprule
		Acronym & Meaning  \\ \\[-1.1em]
		\hline
		CWT      & continuous wavelet transform                    \\ 
		GDW      & Gaussian-derived wavelet                        \\ 
		CW-GDW   & cosine-weighted Gaussian-derived wavelet        \\ 
		WPS      & wavelet power spectrum                          \\ 
		WCC      & wavelet cross-correlation                       \\ 
		env-WPS  & environment-dependent wavelet power spectrum    \\ 
		env-WCC  & environment-dependent wavelet cross-correlation \\ 
		env-bias & environment-dependent bias function             \\ 
		\hline
	\end{tabular}
\end{table}

\section{Data sets}
\label{sec:data}

For our analysis, we utilize three state-of-the-art cosmological simulations: Illustris\footnote{\url{https://www.illustris-project.org/}} \citep{Vogelsberger2013, Vogelsberger2014, Genel2014, Nelson2015}, IllustrisTNG\footnote{\url{https://www.tng-project.org/}} \citep{Marinacci2018, Naiman2018, Nelson2018, Nelson2019, Pillepich2018b, Springel2018}, and SIMBA\footnote{\url{http://simba.roe.ac.uk/}} \citep{Dave2019}. In particular, we mainly focus on the density fields drawn from the \texttt{IllustrisTNG100-1} run at redshifts $z=0$, $1$, $2$, and $3$. The present-day ($z=0$) density fields of the \texttt{Illustris-1} and fiducial SIMBA run (\texttt{m100n1024}) are used for comparison to that of the \texttt{IllustrisTNG100-1}.

\begin{enumerate}[label=(\roman*),wide, labelwidth=!, labelindent=0pt]
    \item The Illustris simulation tracks the evolution of the baryonic and dark matter from redshift $z=127$ to the present day, by using the adaptive moving-mesh AREPO code \citep{Springel2010}. The adopted cosmological parameters are $\Omega_\Lambda = 0.7274$, $\Omega_m=\Omega_\mathrm{dm}+\Omega_\mathrm{b}=0.2726$, $\Omega_\mathrm{b}=0.0456$, $\sigma_8=0.809$, $n_s=0.9631$, and $h=0.704$, consistent with the constraints from WMAP-9 \citep{Bennett2013}. The subgrid models of galaxy formation are described in detail in \citet{Vogelsberger2013} and \citet{Torrey2014}, including star formation and associated supernova feedback, supermassive black hole accretion, and related AGN feedback (thermal-mode and kinetic-mode). In fact, the AGN feedback in the Illustris is so effective that it can lead to a lower gas fraction in the halos, which is in conflict with the observations \citep{Genel2014}. The \texttt{Illustris-1} run that we used has a box of volume $(75 \ h^{-1}\mathrm{Mpc})^3 \simeq (106.5 \ \mathrm{Mpc})^3$ and contains $1820^3$ dark matter particles of mass $\sim 6.26\times 10^6 \ M_\odot$ plus $1820^3$ initial gas cells with average mass $\sim 1.26\times 10^6 \ M_\odot$.
    
    \item The IllustrisTNG (hereafter, TNG) simulation is the successor of the Illustris project, which is also executed with the AREPO code and consists of three simulation volumes: \texttt{TNG300}, \texttt{TNG100} and \texttt{TNG50}. All these runs assume Planck concordance cosmology \citep{Planck2016}, i.e. $\Omega_\Lambda=0.6911$, $\Omega_m=\Omega_\mathrm{dm} + \Omega_\mathrm{b} = 0.3089$, $\Omega_\mathrm{b}=0.0486$, $\sigma_8=0.8159$, $n_s=0.9667$, and $h=0.6774$. Compared with earlier Illustris simulations, TNG uses updated physical models and numerical methods and yields better consistency with the available observations of galaxy formation and evolution. Thus, it provides us with an ideal laboratory for testing theoretical models and developing new analysis tools to achieve a more precise understanding of the structure formation and galaxy formation. The details of the TNG model for galaxy formation are given in \citet{Pillepich2018a} and \citet{Weinberger2018}. In this work, we use density fields of the \texttt{TNG100-1} simulation at redshifts $z=0.0$, $1.0$, $2.0$, and $3.0$. This simulation adopts a periodic box of side length $L_\mathrm{box}=75 \ h^{-1}\mathrm{Mpc}\simeq 110.7 \ \mathrm{Mpc}$, and uses the same initial conditions as \texttt{Illustris-1}. The mass of each dark matter particle  is $\sim 7.5\times10^6 \ M_\odot$ and the initial baryonic mass resolution is $\sim 1.4 \times 10^6 \ M_\odot$. 
    
    \item The SIMBA simulation suite is performed with the meshless finite-mass hydrodynamics code GIZMO \citep{Hopkins2015}, and includes four simulation volumes: \texttt{m100n1024}, \texttt{m50n1024}, \texttt{m25n1024} and \texttt{m12.5n1024}, all starting at $z=249$ and ending at $z=0$. These simulations also assume the Planck concordance cosmology \citep{Planck2016}: $\Omega_\Lambda=0.7$, $\Omega_m=\Omega_\mathrm{dm} + \Omega_\mathrm{b} = 0.3$, $\Omega_\mathrm{b}=0.048$, $\sigma_8=0.82$, $n_s=0.97$, $h=0.68$. The implementation of the galaxy formation physics in this simulation, e.g. AGN feedback, stellar feedback, gas cooling, and star formation, can be found in \citet{Dave2019}. With these sophisticated models, SIMBA is also capable of reproducing a wide range of observations. The fiducial SIMBA run (\texttt{m100n1024}) used here has a box size of $100 \ h^{-1}\mathrm{Mpc} \simeq 147 \ \mathrm{Mpc}$ with $1024^3$ dark matter particles plus $1024^3$ gas cells. The mass resolution is $\sim 9.6\times 10^7 \ M_\odot$ for dark matter and $\sim 1.82\times 10^7 \ M_\odot$ for gas.
\end{enumerate}
\begin{figure}[t]
	\centerline{\includegraphics[width=0.45\textwidth]{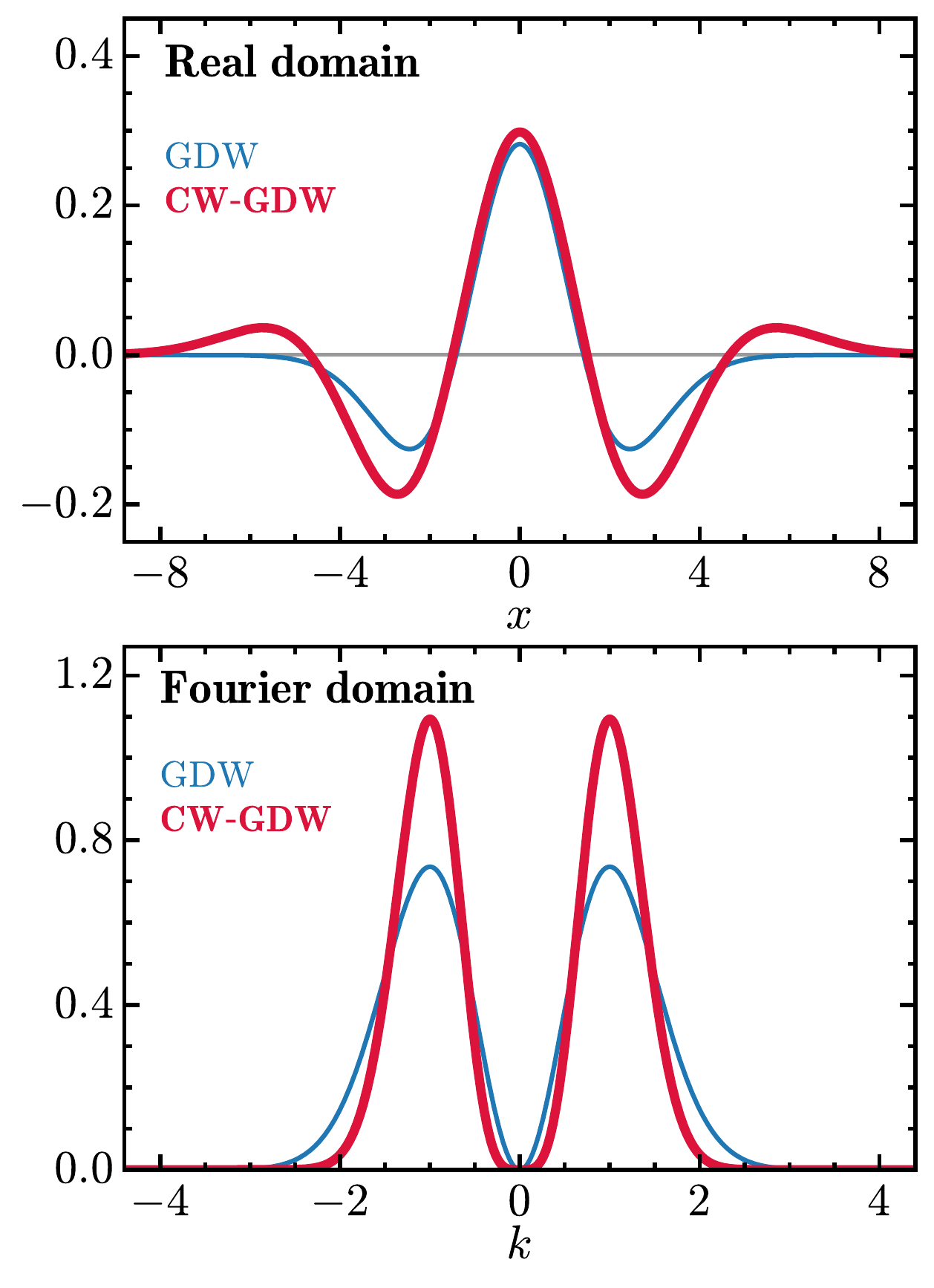}}
	\caption{Comparison of the 1D GDW (thin line) and CW-GDW (thick line) in the real (top) and Fourier (bottom) domain at scale $w=1$. For presentational convenience, all variables are dimensionless.}
	\label{fig:CW-GDW_1d}
\end{figure}

\section{Continuous wavelet methods}
\label{sec:cwt_methods}

\subsection{The 1D Cosine-weighted Gaussian-derived Wavelet}
\label{sec:CW-GDW_1d}

In previous works \citep{Wang2021a, Wang2021b}, we applied the GDW, a low-oscillation wavelet, to the spectral analysis of the matter clustering in 1D and isotropic cases. However, low-oscillation wavelets are more extended in Fourier space, which can mean that the results of spectral analysis at small scales are contaminated by large scales \citep{Frick2001}. To achieve a better separation of scales, we take the Gaussian function weighted by the cosine as the smoothing function, then derive the wavelet from it. In 1D, the form of such a cosine-weighted Gaussian-derived wavelet (CW-GDW) is constructed as below. From the cosine-weighted Gaussian smoothing function
\begin{flalign*}
g_\mathrm{cos}(w,x)=\frac{w}{\sqrt{2\pi \alpha^2}}\exp\left( \frac{1-(w x/\alpha)^2}{2} \right)\cos\left(\frac{w x}{\alpha} \right), &&
\end{flalign*}
with its Fourier transform
\begin{flalign*}
\hat g_\mathrm{cos}(w,k)=\cosh\left( \frac{\alpha k}{w} \right)\exp\left(-\frac{(\alpha k/w)^2}{2}\right), &&
\end{flalign*}
we have the CW-GDW
\begin{flalign}
\label{eq:CW-GDW_1d}
\psi(w,x) &\equiv w^\kappa\frac{\partial g_\mathrm{cos}(w,x)}{\partial w} \nonumber\\
& = \frac{w^\kappa}{\sqrt{ 2\pi \alpha^2 }}\Biggl[ \left(1-\left(\frac{wx}{\alpha}\right)^2\right)\cos\left(\frac{wx}{\alpha} \right) \nonumber\\
& \quad- \left(\frac{wx}{\alpha} \right)\sin\left(\frac{wx}{\alpha} \right) \Biggr]\exp\left(\frac{1-(\frac{wx}{\alpha})^2}{2}\right), &&
\end{flalign}
and its Fourier transform
\begin{flalign}
\label{eq:CW-GDW_k_1d}
\hat \psi(w,k) & = w^{-\kappa}\left(\frac{\alpha k}{w}\right)\Biggl[\left(\frac{\alpha k}{w}\right)\cosh\left(\frac{\alpha k}{w}\right) \nonumber\\
& \quad - \sinh\left(\frac{\alpha k}{w}\right)\Biggr]\exp\left(-\frac{1}{2}\left(\frac{\alpha k}{w}\right)^2\right), &&
\end{flalign}
where the dimensionless constant $\alpha \simeq 2.20473$, which is used to set $w$ to be the peak frequency of $\hat\psi(w,k)$. The index $\kappa = 1/2$ guarantees that the integral of the square of the wavelet is conserved across different scales. Compared to the GDW, the CW-GDW has a better spectral resolution, as shown in Fig.~\ref{fig:CW-GDW_1d}. This wavelet can be generalized to higher dimensions, for analyzing the large-scale structures of the universe. The isotropic and anisotropic CW-GDWs are both described as follows.
\begin{figure}[t]
	\centerline{\includegraphics[width=0.46\textwidth]{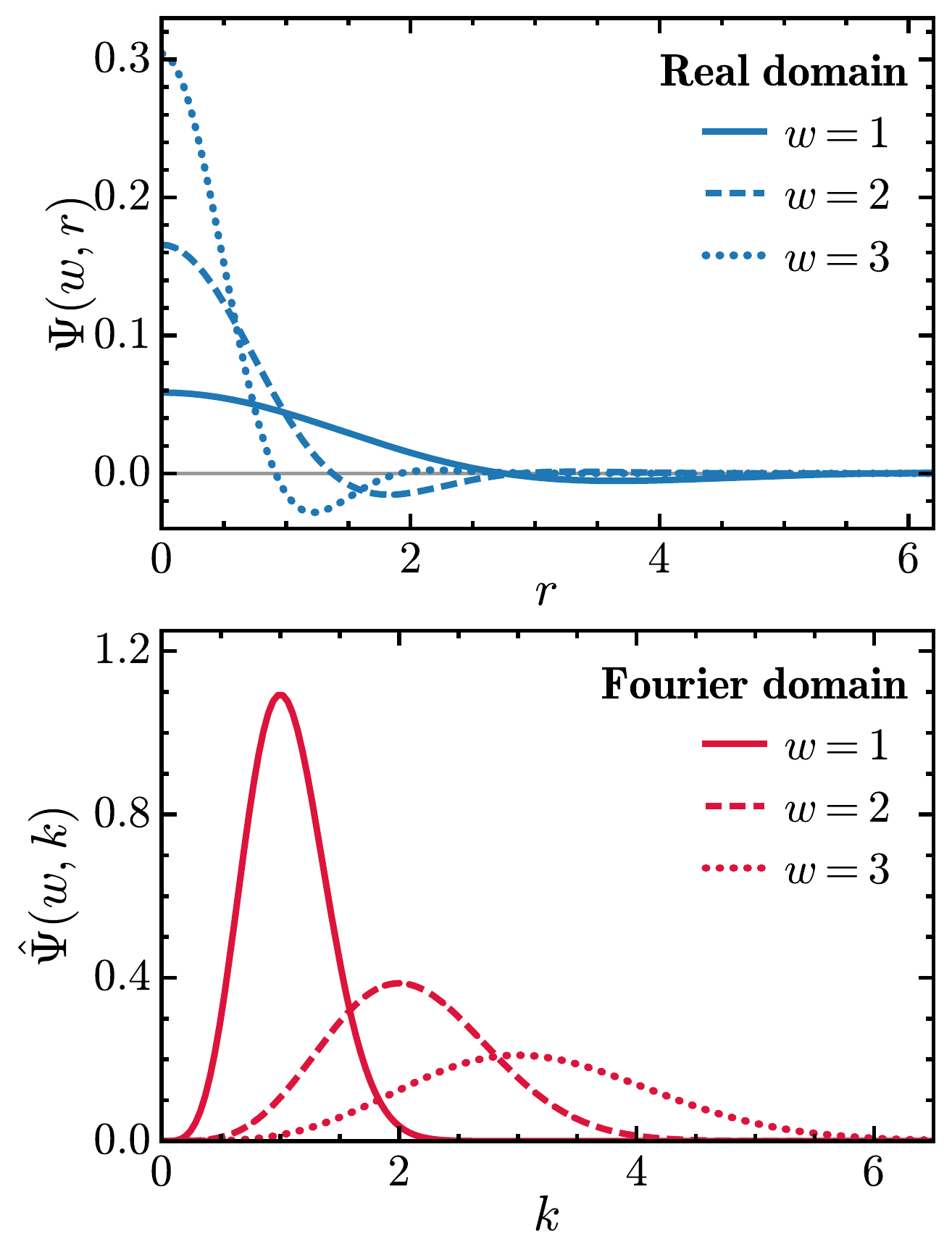}}
	\caption{The isotropic CW-GDW in the real (top) and Fourier (bottom) domain at three scales, $w=1$, $2$, and $3$. For presentational convenience, all variables are dimensionless.}
	\label{fig:CW-GDW_iso}
\end{figure}
\begin{figure*}[t]
	\centering
	\subfigure{\includegraphics[width=0.33\textwidth]{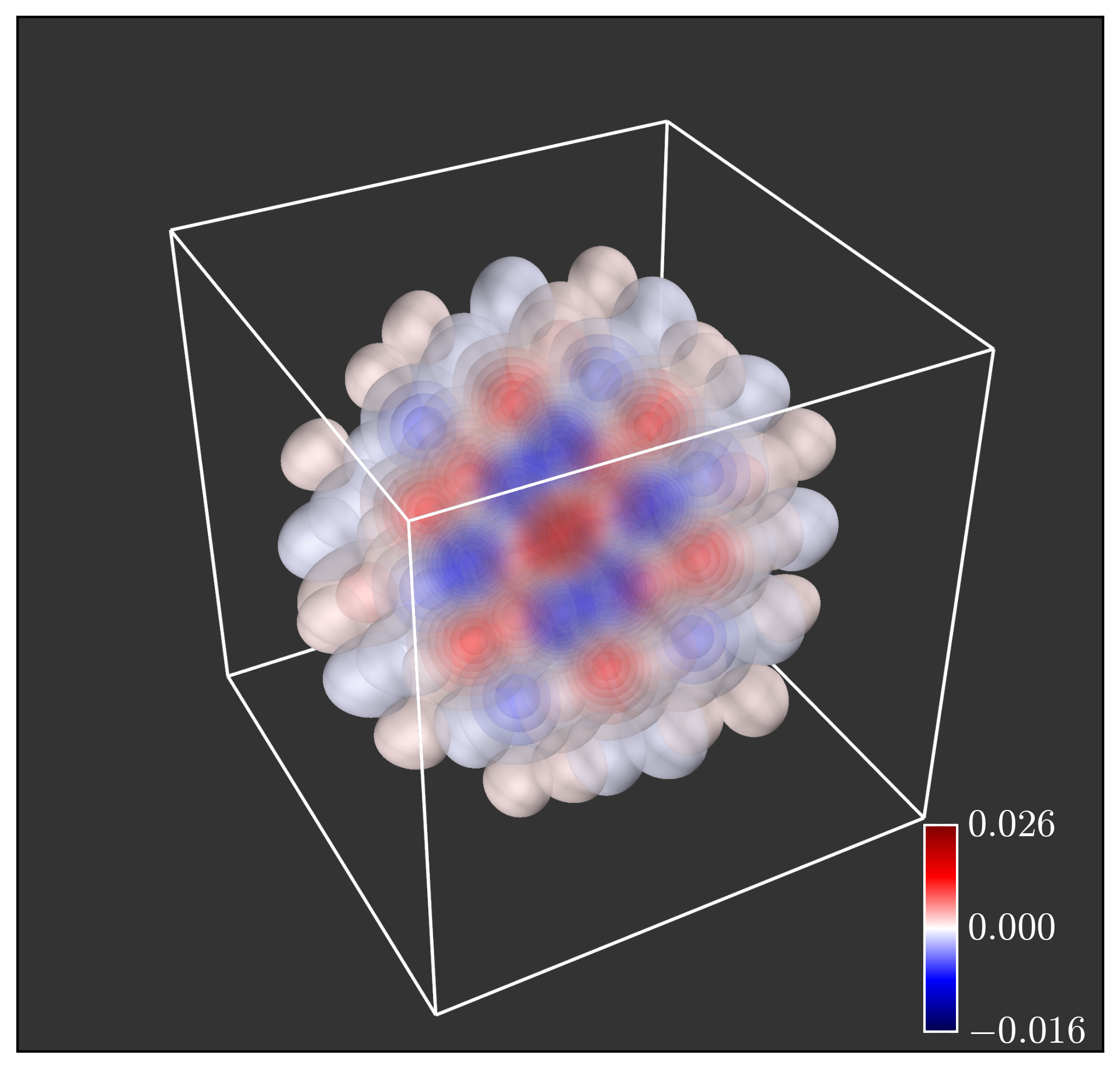}}
	\subfigure{\includegraphics[width=0.33\textwidth]{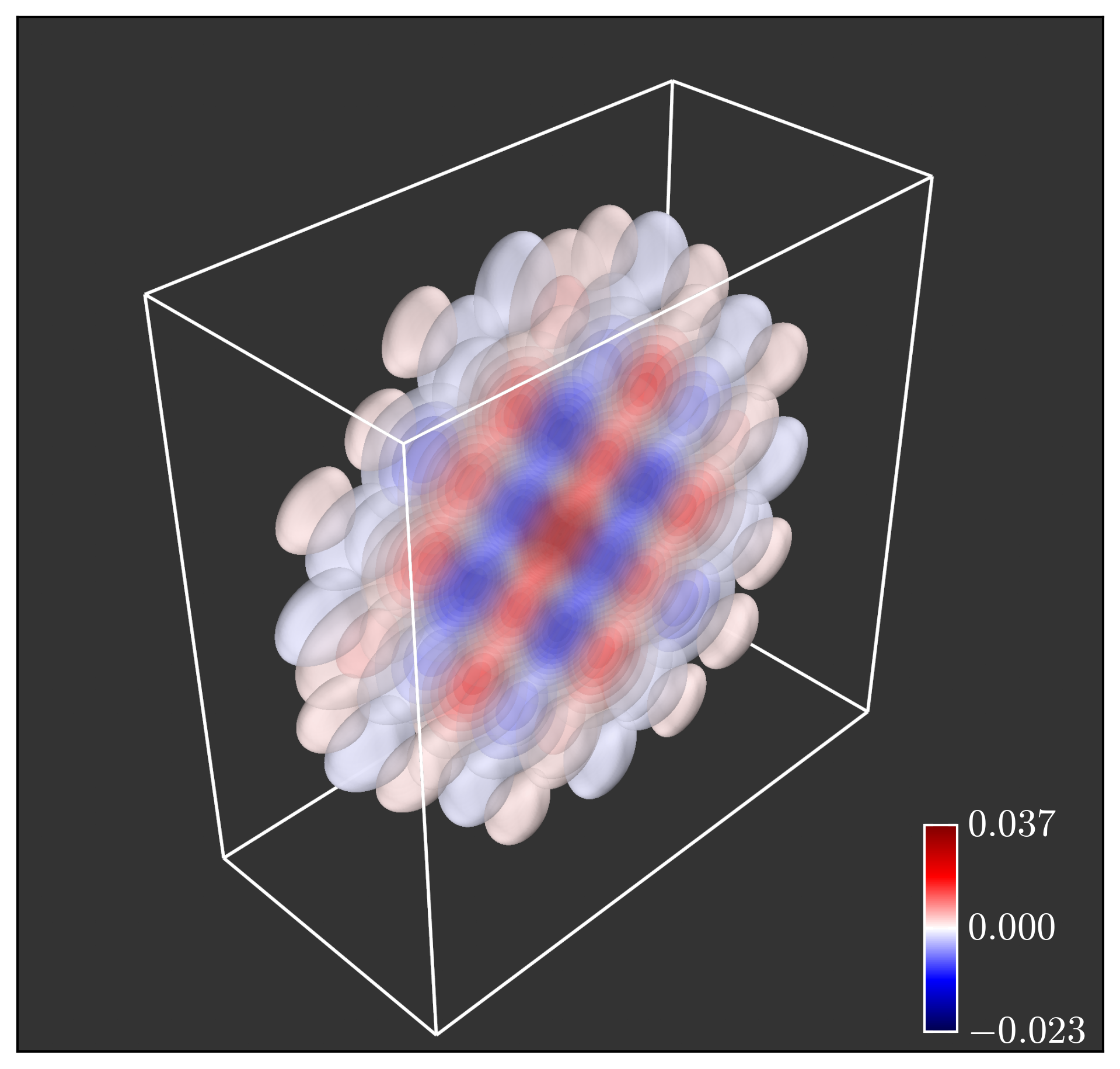}}
	\subfigure{\includegraphics[width=0.33\textwidth]{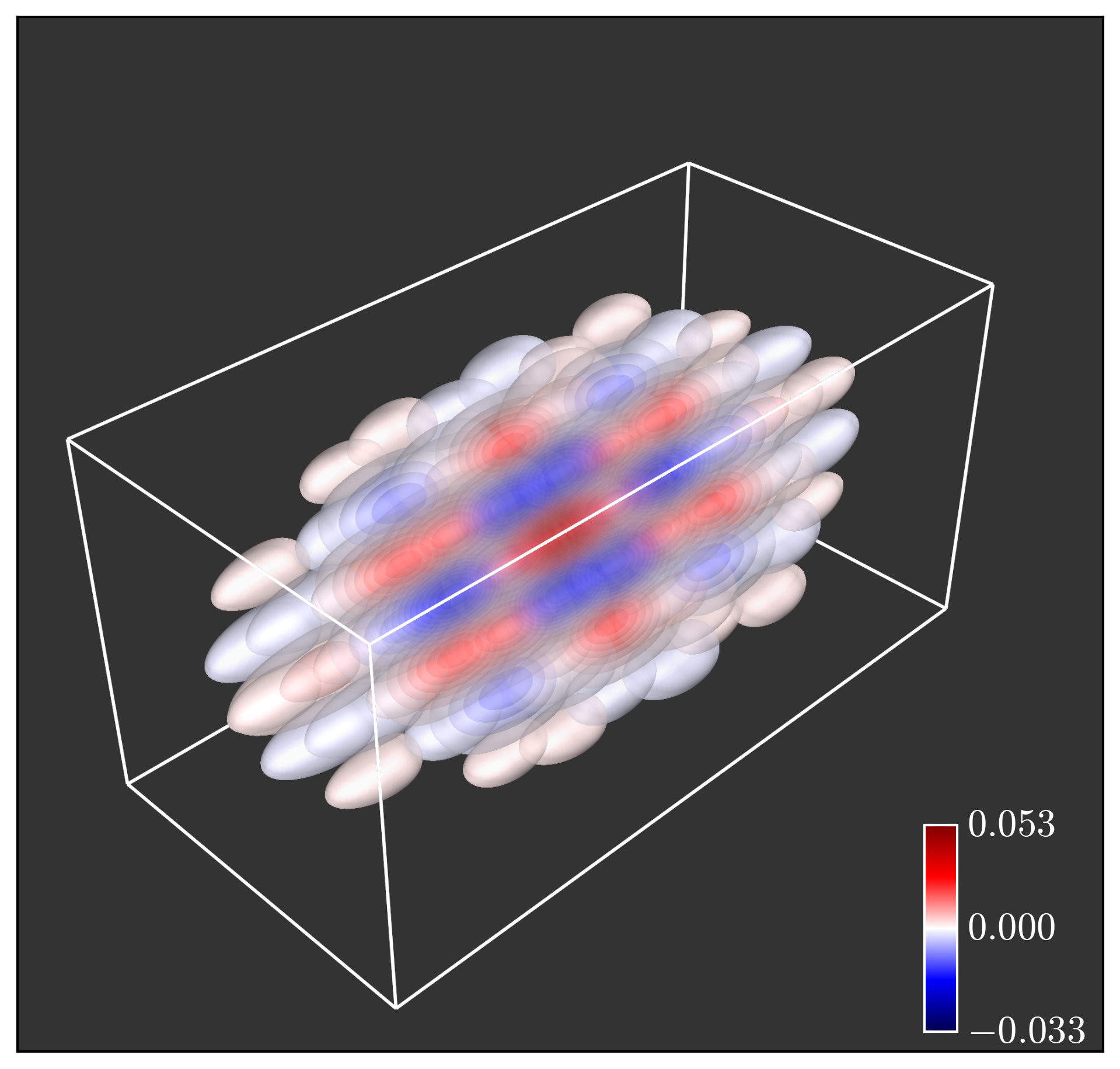}}
	\subfigure{\includegraphics[width=0.33\textwidth]{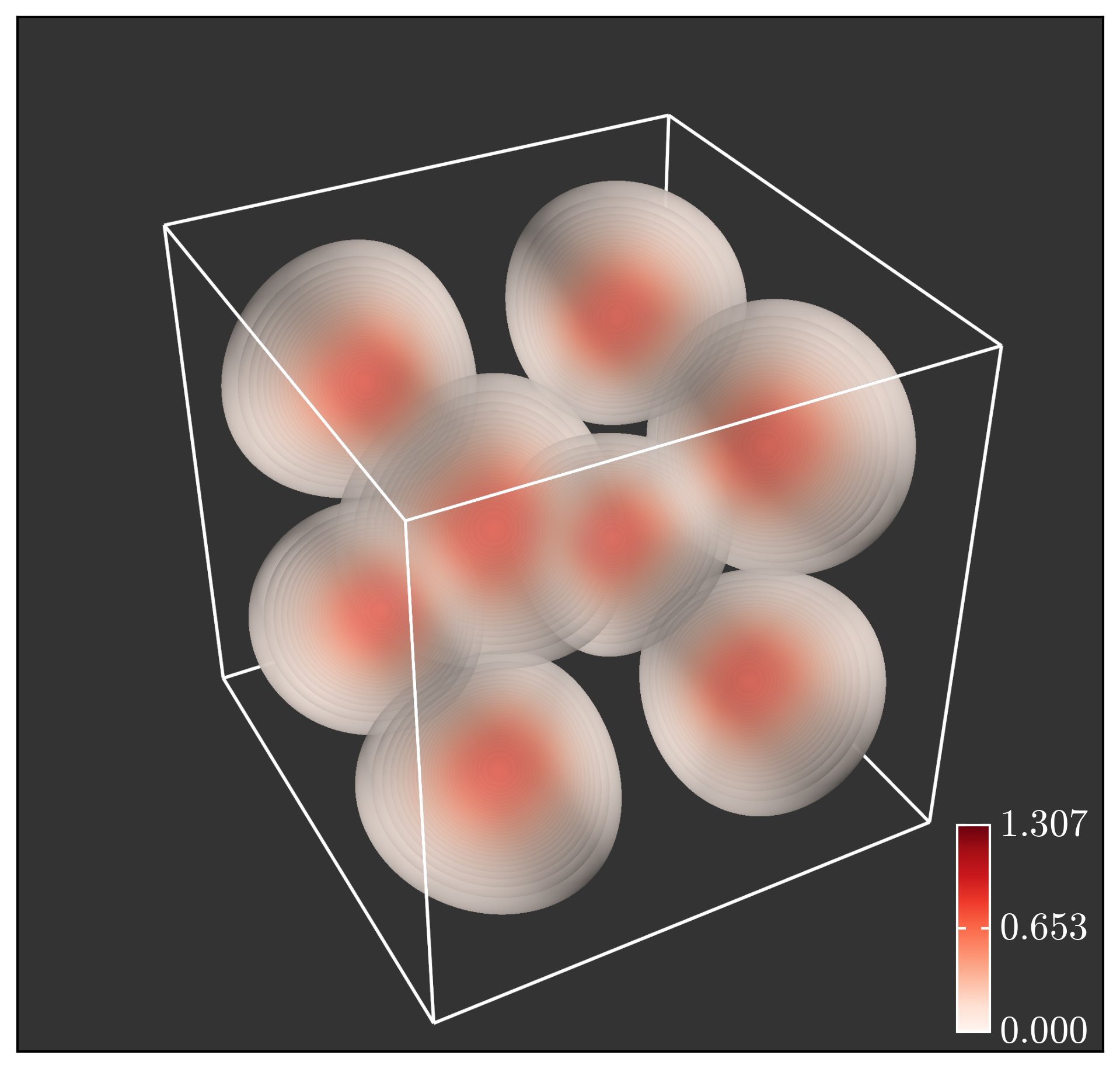}}
	\subfigure{\includegraphics[width=0.33\textwidth]{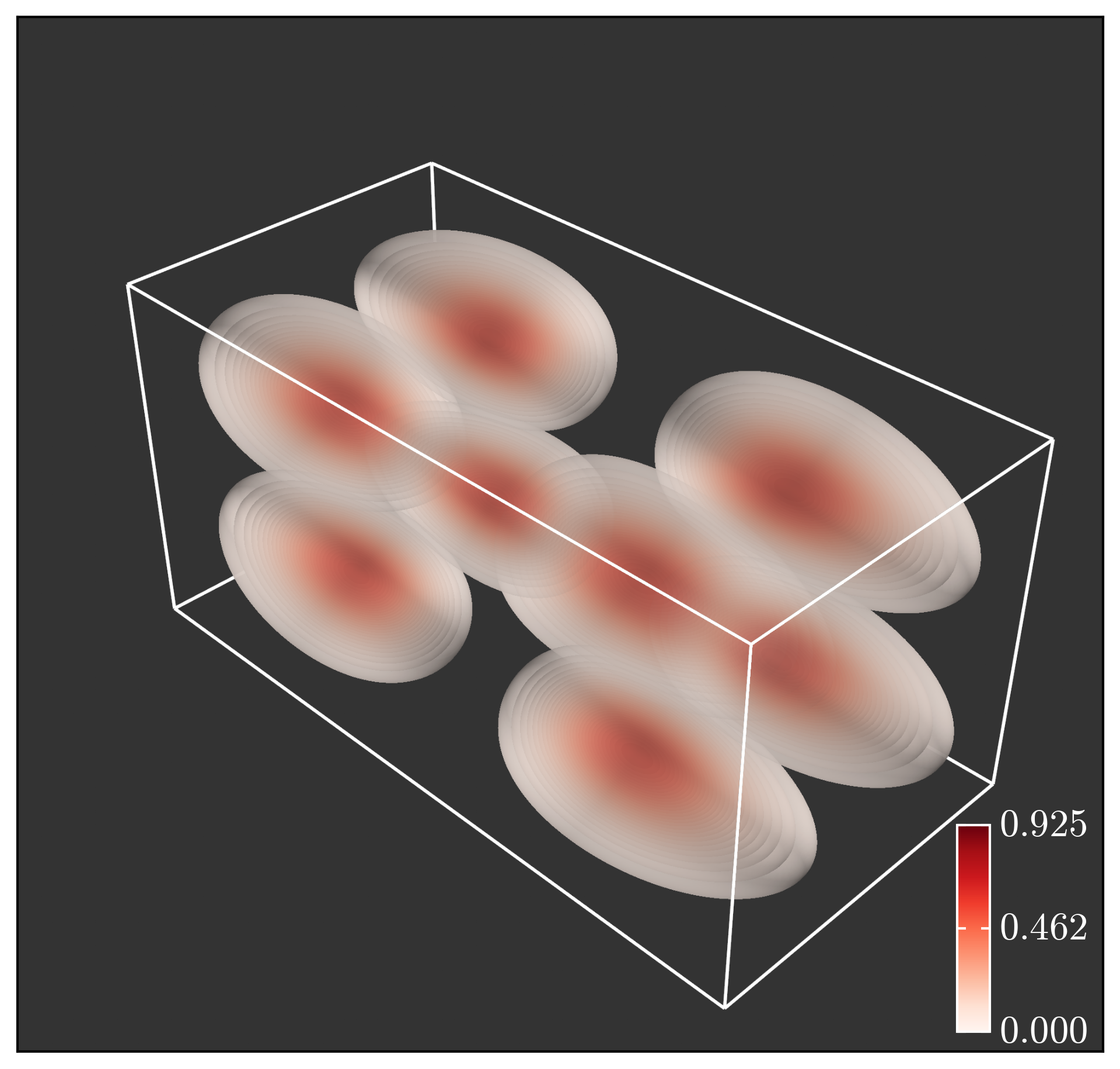}}
	\subfigure{\includegraphics[width=0.33\textwidth]{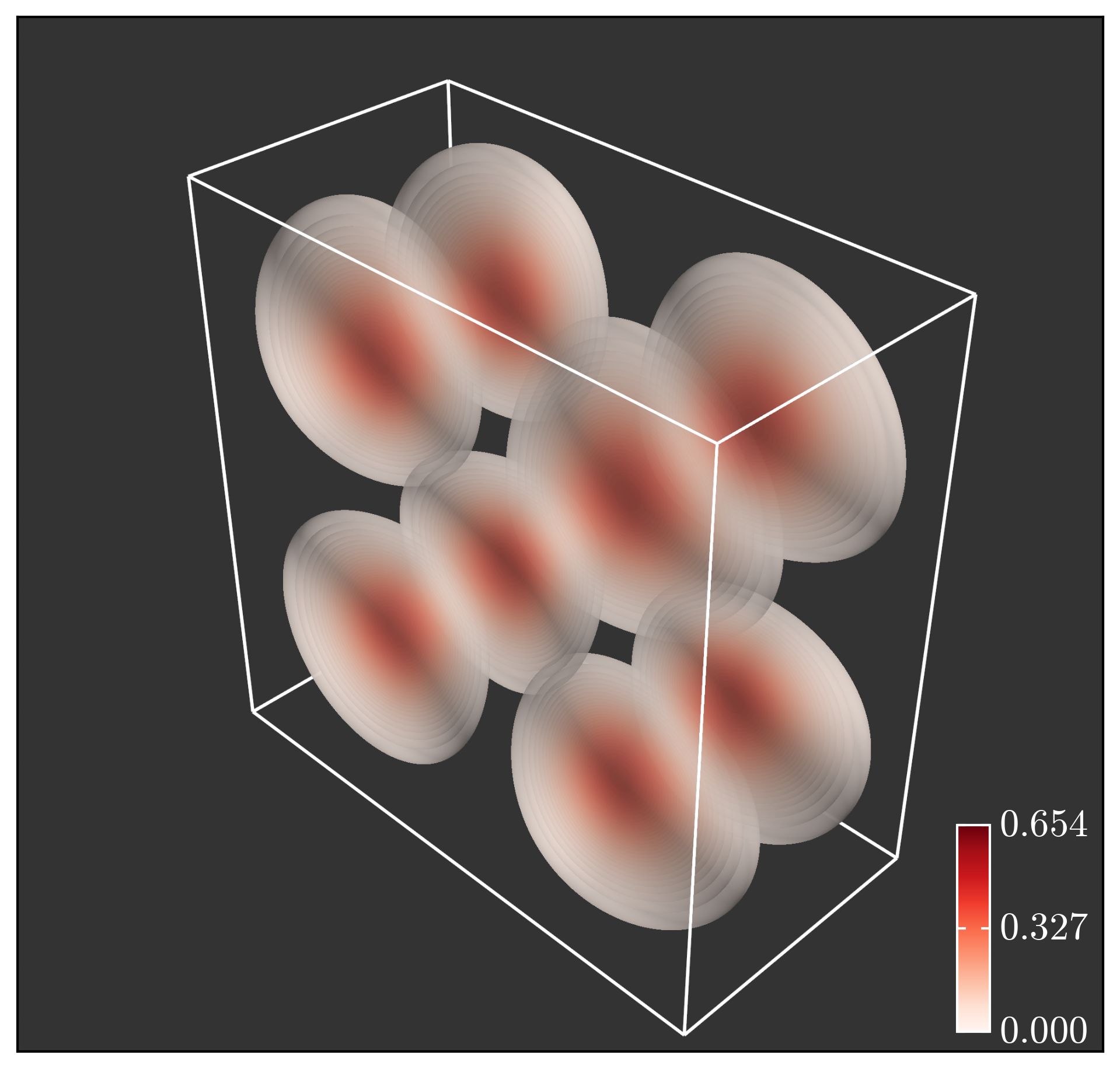}}
	\caption{Contour plots of the anisotropic CW-GDW with different scales in real space (upper panels) and in Fourier space (lower panels). \textit{Left column}: the anisotropic CW-GDW and its Fourier counterpart at scale $\mathbf{w}=(1,1,1)$. \textit{Middle column}: the anisotropic CW-GDW and its Fourier counterpart at scale $\mathbf{w}=(1,2,1)$. \textit{Right column}: the anisotropic CW-GDW and its Fourier counterpart at scale $\mathbf{w}=(1,2,2)$.}
	\label{fig:CW-GDW_aniso}
\end{figure*}

\begin{figure*}[t]
	\centering
	\subfigure{\includegraphics[width=0.78\textwidth]{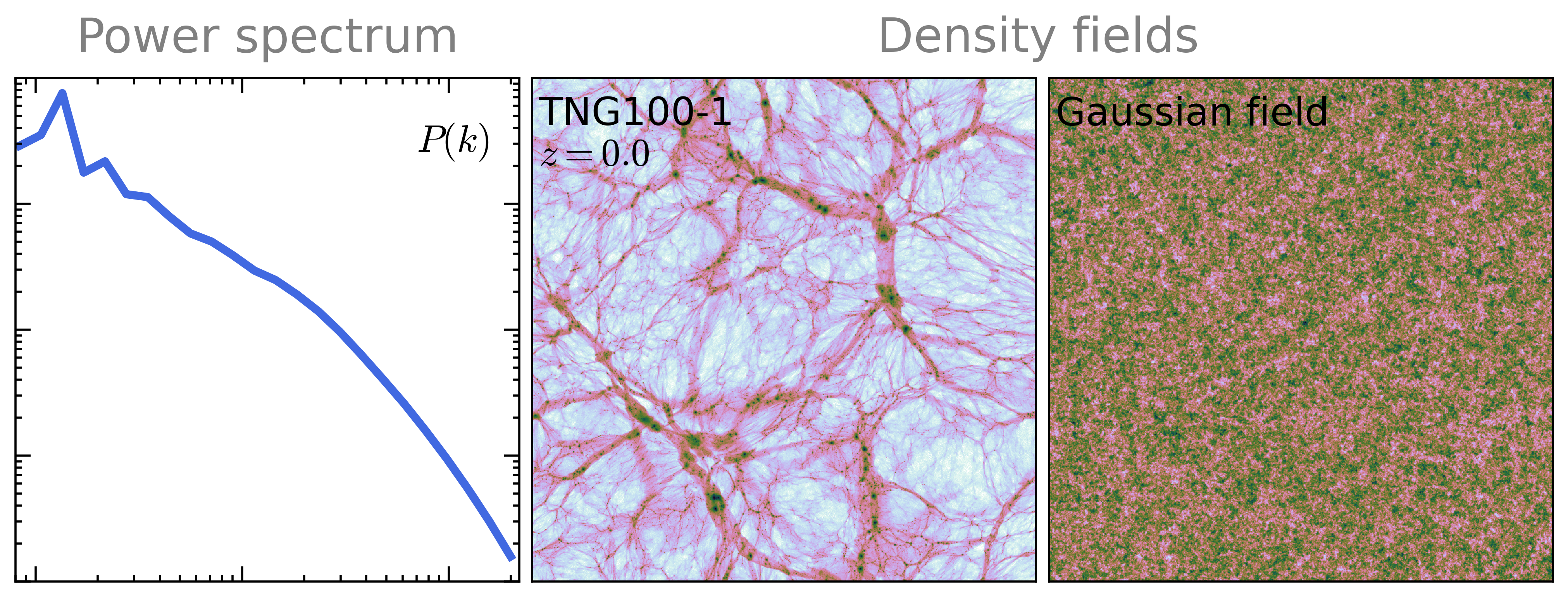}}
	\subfigure{\includegraphics[width=0.98\textwidth]{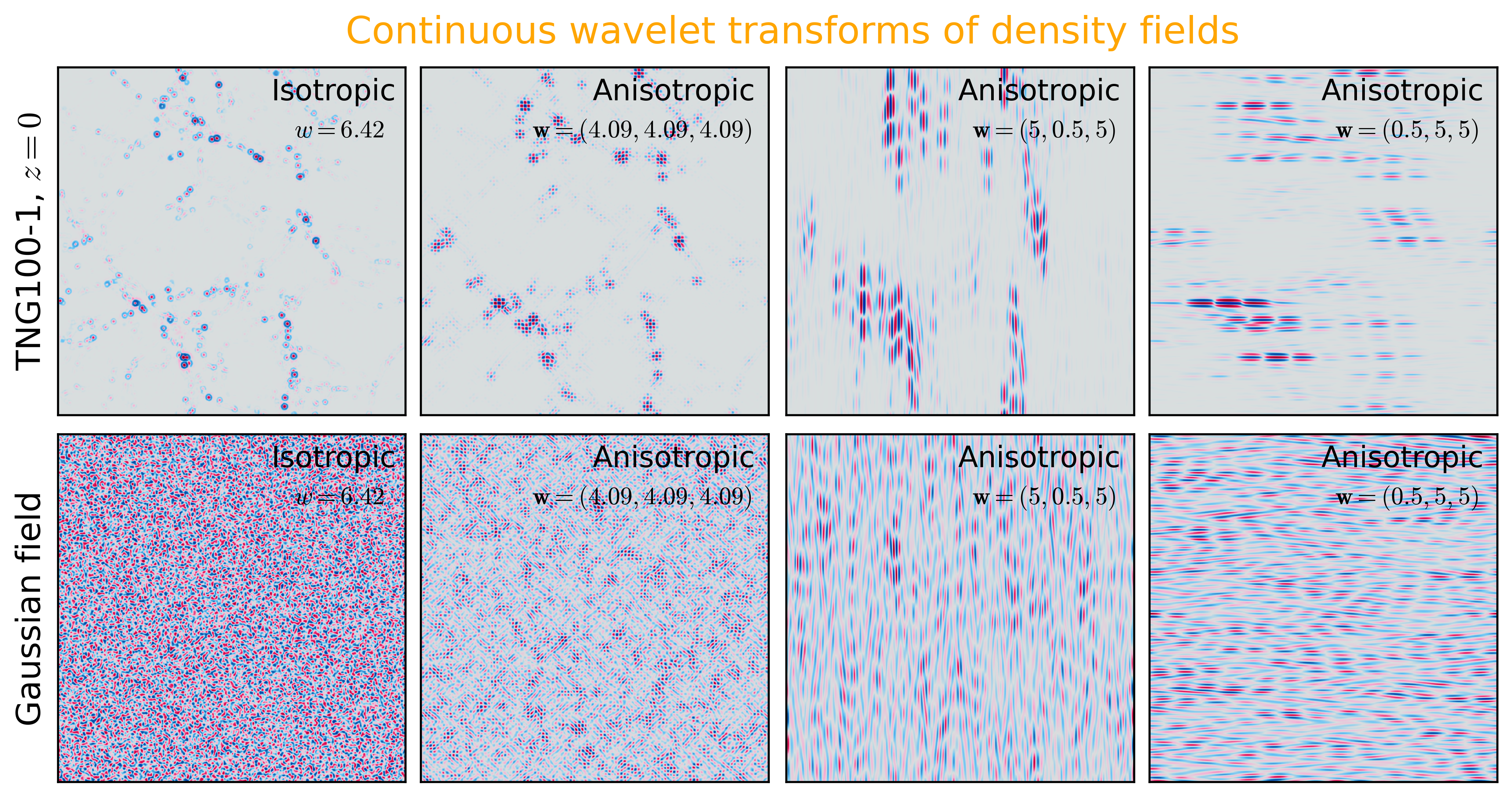}}
	\caption{\textit{Top row}: the \texttt{TNG100-1} dark matter density field at $z=0$ and the Gaussian field with the same Fourier power spectrum in a $75\times 75 \ h^{-2}\mathrm{Mpc}^2$ slice of thickness $1 \ h^{-1}\mathrm{Mpc}$. \textit{Middle row}: the CWTs of the \texttt{TNG100-1} field in the same slice. Specifically, the left plot is isotropic CW-GDW transform at scale $w=6.42 \ h\mathrm{Mpc}^{-1}$. The three plots on the right are anisotropic CW-GDW transforms at scales of $\mathbf{w}=(4.09,4.09,4.09) \ h\mathrm{Mpc}^{-1}$, $(5,0.5,5) \ h\mathrm{Mpc}^{-1}$ and $(0.5,5,5) \ h\mathrm{Mpc}^{-1}$, respectively. \textit{Bottom row}: the same as the middle row, but for the Gaussian field with the equivalent power spectrum. By using equations~\eqref{eq:scale_wavenumber_iso} and \eqref{eq:scale_wavenumber_aniso}, it can be seen that all these wavelet scales correspond to a Fourier scale of $|\mathbf{k}| \approx 7.5 \ h\mathrm{Mpc}^{-1}$.}
	\label{fig:3d_cwts}
\end{figure*}

\subsection{The 3D Isotropic CW-GDW Transform}
\label{sec:CW-GDW_iso}

The 3D isotropic CW-GDW is defined in Fourier space by taking the derivative of the isotropic-cosine weighted Gaussian smoothing function $\hat G_\mathrm{cos}(w,\mathbf{k})$ w.r.t. $w$,

\begin{flalign}
\label{eq:CW-GDW_iso}
\hat\Psi(w,\mathbf{k}) &\equiv w^\kappa \frac{\partial \hat G_\mathrm{cos}(w,\mathbf{k})}{\partial w} \nonumber \\
& = w^{\kappa-1}\left( \frac{\alpha |\mathbf{k}|}{w} \right) \Bigg[ \left( \frac{\alpha |\mathbf{k}|}{w} \right)\cosh\left( \frac{\alpha |\mathbf{k}|}{w} \right) \nonumber \\
& \quad - \sinh\left( \frac{\alpha |\mathbf{k}|}{w} \right) \Bigg]\exp\left(-\frac{\alpha^2 |\mathbf{k}|^2}{2w^2}  \right), &&
\end{flalign}
with
\begin{flalign*}
\hat G_\mathrm{cos}(w,\mathbf{k}) = \cosh\left( \frac{\alpha |\mathbf{k}|}{w} \right)\exp\left(-\frac{\alpha^2 |\mathbf{k}|^2}{2w^2}  \right), &&
\end{flalign*}
where the index $\kappa=-1/2$, in this case. It can be seen that $\hat\Psi(w,\mathbf{k})$ depends only on the magnitude of the wavevector, i.e. $\hat\Psi(w,\mathbf{k})=\hat\Psi(w,k)$, which is shown in the lower panel of Fig. \ref{fig:CW-GDW_iso}. The isotropic CW-GDW $\Psi(w,\mathbf{r})$ in real space can be computed numerically by FFT, which is plotted in the upper panel of Fig. \ref{fig:CW-GDW_iso}. 

By convolving the density contrast $\delta(\mathbf{r})=(\rho(\mathbf{r})-\overline{\rho})/\overline{\rho}$ with the isotropic CW-GDW, we get the wavelet transform
\begin{flalign}
\label{eq:CWT_iso}
W(w,\mathbf{r}) &= \iiint\limits_{-\infty}^{+\infty} \delta(\mathbf{u})\Psi(w,\mathbf{r}-\mathbf{u})\mathrm{d}^3\mathbf{u} \nonumber \\
& = \frac{1}{(2\pi)^3}\iiint\limits_{-\infty}^{+\infty} \hat\delta(\mathbf{k})\hat\Psi(w,\mathbf{k})e^{-i\mathbf{k}\cdot\mathbf{r}}\mathrm{d}^3\mathbf{k}, &&
\end{flalign}
where $\hat\delta(\mathbf{k})$ is the Fourier transform of $\delta(\mathbf{r})$, which is estimated by interpolating the mass of each particle to a regular grid with $1024^3$ cells, using the cloud-in-cell method. From Equation \eqref{eq:CWT_iso}, the CWT corresponds to a local filtering of $\delta(\mathbf{r})$ around wavenumber $|\mathbf{k}|\sim w$. According to Appendix B of \cite{Wang2021b}, the original density field can be reconstructed by 
\begin{flalign}
\label{eq:inverse_CWT_iso}
\delta(\mathbf{r}) = \int_0^{+\infty} w^{-\kappa}W(w,\mathbf{r})\mathrm{d}w. &&
\end{flalign}

In many investigations, the comparison between a wavelet transform and a Fourier transform at a given scale requires the correspondence between the wavelet scale $w$ and the Fourier wavenumber $|\mathbf{k}|$, which is
\begin{flalign}
\label{eq:scale_wavenumber_iso}
w= c_w |\mathbf{k}|, &&
\end{flalign}
with $c_w \simeq 0.85617$ for the isotropic CW-GDW, determined by the wavelet spectral peak of a harmonic wave with frequency $|\mathbf{k}|$ \citep[e.g.][]{Meyers1993, Torrence1998, Wang2021b}.

\subsection{The 3D anisotropic CW-GDW transform}

Following the logic of \citet{Wang2021a}, the 3D anisotropic CW-GDW can be obtained from the 1D version, as shown below:
\begin{flalign}
\label{eq:CW-GDW_aniso}
\Psi(\mathbf{w},\mathbf{r}) & \equiv (w_xw_yw_z)^\kappa\frac{\partial^3G_\mathrm{cos}(\mathbf{w},\mathbf{r})}{\partial w_x\partial w_y\partial w_z} \nonumber\\
& = w_x^\kappa\frac{\partial g_\mathrm{cos}(w_x, x)}{\partial w_x}w_y^\kappa\frac{\partial g_\mathrm{cos}(w_y, y)}{\partial w_y}w_z^\kappa\frac{\partial g_\mathrm{cos}(w_z, z)}{\partial w_z} \nonumber\\
& = \psi(w_x, x)\psi(w_y, y)\psi(w_z,z), &&
\end{flalign}
with the Fourier transform
\begin{flalign}
\label{eq:CW-GDW_aniso_k}
\hat\Psi(\mathbf{w},\mathbf{k}) = \hat\psi(w_x,k_x)\hat\psi(w_y,k_y)\hat\psi(w_z,k_z), &&
\end{flalign}
where the index $\kappa = 1/2$, $G_\mathrm{cos}(\mathbf{w}, \mathbf{r}) = g_\mathrm{cos}(w_x, x) g_\mathrm{cos}(w_y, y)g_\mathrm{cos}(w_z, z)$ is the anisotropic cosine-weighted Gaussian smoothing function with $\mathbf{r}=(x,y,z)$, and $\mathbf{w}=(w_x, w_y,w_z)$ is the scale vector. In contrast to the isotropic CW-GDW, which can only dilate or contract in the radial direction, the anisotropic one can have different dilations along different spatial axes. For example, three scale configurations---$\mathbf{w} = (1, 1, 1)$, $(1, 2, 1)$, and $(1, 2, 2)$---are illustrated in Fig.~\ref{fig:CW-GDW_aniso}. Intuitively, the CW-GDW with a scale of $(1, 1, 1)$ may have a stronger response to the clumps of the cosmic web, while those with scales of $(1, 2, 1)$ and $(1, 2, 2)$ may have better responses to the sheet and filamentary structures, respectively.

By using the anisotropic CW-GDW, the wavelet transform of the density contrast $\delta(\mathbf{r})$ is given by
\begin{flalign}
\label{eq:CWT_aniso}
W(\mathbf{w},\mathbf{r}) &= \iiint\limits_{-\infty}^{+\infty} \delta(\mathbf{u})\Psi(\mathbf{w},\mathbf{r}-\mathbf{u})\mathrm{d}^3\mathbf{u} \nonumber \\
& = \frac{1}{(2\pi)^3}\iiint\limits_{-\infty}^{+\infty} \hat\delta(\mathbf{k})\hat\Psi(\mathbf{w},\mathbf{k})e^{-i\mathbf{k}\cdot\mathbf{r}}\mathrm{d}^3\mathbf{k}. &&
\end{flalign}
In this anisotropic case, the reconstruction formula is a triple integral w.r.t. the scale space,
\begin{flalign}
\label{eq:inverse_CWT_aniso}
\delta(\mathbf{r})=\iiint\limits_0^{+\infty} w_x^{-\kappa} w_y^{-\kappa} w_z^{-\kappa} W(\mathbf{w}, \mathbf{r}) \mathrm{d}^3\mathbf{w}, &&
\end{flalign}
which is the extension of the 1D reconstruction formula given in \citet{Wang2021b}. The relation between the wavelet scale $\mathbf{w}=(w_x,w_y,w_z)$ and the Fourier wavevector $\mathbf{k}=(k_x,k_y,k_z)$ is
\begin{flalign}
\label{eq:scale_wavenumber_aniso}
w_x = c_w |k_x|, \ \
w_y = c_w |k_y|, \ \
w_z = c_w |k_z|, &&
\end{flalign} 
with $c_w \simeq 0.94411$, determined by the wavelet spectral peak of a harmonic wave in each spatial axis.

\subsection{Wavelet statistics}
\label{sec:measure_procedure}

Being the inverse Fourier transforms of $\hat\delta(\mathbf{k})\hat\Psi(w,\mathbf{k})$ or $\hat\delta(\mathbf{k})\hat\Psi(\mathbf{w},\mathbf{k})$, the CWTs of density fields can be efficiently implemented by FFT. For illustration, we compare in Fig.~\ref{fig:3d_cwts} the CWTs of the \texttt{TNG100-1} dark matter density field $\delta_\mathrm{dm}(\mathbf{r})$ at $z=0$ and the Gaussian field constructed by randomizing phases of $\hat\delta_\mathrm{dm}(\mathbf{k})$. Therefore, these two fields share the same power spectrum, which is totally insensitive to the non-Gaussianity. In contrast, the CWT of the non-Gaussian dark matter field is completely different from that of the Gaussian field, which is attributed to the CWT preserving both the scale information and the spatial texture of the fields. Notice that clusters are highlighted by the isotropic CW-GDW and anisotropic CW-GDW, with a scale of $\mathbf{w}=(4.09,4.09,4.09) \ h\mathrm{Mpc}^{-1}$, and the vertical and horizontal filaments are captured by the anisotropic CW-GDWs with $\mathbf{w}=(5,0.5,5) \ h\mathrm{Mpc}^{-1}$ and $\mathbf{w}=(0.5,5,5) \ h\mathrm{Mpc}^{-1}$, respectively. Because the CWT of the density field varies with the spatial coordinate $\mathbf{r}$ at a given scale, it should be a function of the local density fluctuation $\delta(\mathbf{r})$. With this in mind, we can interpret the simultaneous dependence of the matter clustering on the scale and the density environment using statistics developed from the CWT.

We now introduce some wavelet statistics by means of the isotropic CW-GDW transform. To measure the clustering strength in different environments, we take the average of the wavelet coefficients at each scale with the same local density, i.e., $\delta(\mathbf{r})=\delta$, which is designated as the env-WPS, as shown below:
\begin{flalign}
\label{eq:env_wps_iso}
P^W_\mathrm{i}(w,\delta)\equiv\Big\langle |W_\mathrm{i}(w,\mathbf{r})|^2 \Big\rangle_{\delta(\mathbf{r})=\delta}, &&
\end{flalign}
where the subscript `i' refers to either dark matter or gas, and $W_\mathrm{i}(w,\mathbf{r})$ is the isotropic CW-GDW transform of the corresponding density field. We use $\Delta(\mathbf{r}) = \rho(\mathbf{r})/\bar{\rho} = 1 + \delta(\mathbf{r})$ to denote the density field of the total matter, with $\delta(\mathbf{r}) = (\Omega_\mathrm{dm} \delta_\mathrm{dm} (\mathbf{r}) + \Omega_\mathrm{b} \delta_\mathrm{gas}(\mathbf{r}))/\Omega_\mathrm{m}$\footnote{We simply use the gas overdensity here to refer to the total baryon overdensity, since most of the baryons are in gaseous form \citep{Bregman2007}.}. Practically, the densities are divided into 16 bins: (a) 14 logarithmic bins equally divided between $\Delta = 0.1$ and $\Delta = 200$; (b) one with $\Delta < 0.1$ (extreme underdense environments); and (c) one with $\Delta > 200$ (usually the halo environments). If we average over all the possible densities, then the env-WPS will degenerate to the global WPS,
\begin{flalign*}
P^W_\mathrm{i}(w) =\Big\langle |W_\mathrm{i}(w,\mathbf{r})|^2 \Big\rangle_{\mathrm{all} \ \delta(\mathbf{r})} = \Big\langle |W_\mathrm{i}(w,\mathbf{r})|^2 \Big\rangle_{V_\mathrm{box}}, &&
\end{flalign*}
which is proportional to the Fourier power spectrum \citep{Wang2021b}.

We define the env-bias as the square root of the ratio of the env-WPS of gas to that of dark matter,
\begin{flalign}
\label{eq:env_bias_iso}
b^W(w,\delta)\equiv\sqrt{\frac{P^W_\mathrm{gas}(w,\delta)}{P^W_\mathrm{dm}(w,\delta)} }, &&
\end{flalign}
which enables the dark matter CWT to be reconstructed from the gas CWT, as illustrated below:
\begin{flalign*}
W'_\mathrm{dm}(w,\mathbf{r}) = W_\mathrm{gas}(w,\mathbf{r})/b^W(w,\delta), &&
\end{flalign*}
where the gas CWT $W_\mathrm{gas}(w,\mathbf{r})$ is divided by the env-bias $b^W(w,\delta)$ if $\delta(\mathbf{r})$ is in density bin $\delta$. Inspired by the bias research \citep{Bonoli2009} based on the Fourier power spectrum, the reconstruction error can be estimated by
\begin{flalign*}
\epsilon(w,\delta) = \frac{ \langle|W'_\mathrm{dm}(w,\mathbf{r})-W_\mathrm{dm}(w,\mathbf{r})|^2\rangle_{\delta(\mathbf{r})=\delta} }{ \langle |W_\mathrm{dm}(w,\mathbf{r})|^2 \rangle_{\delta(\mathbf{r})=\delta} }. &&
\end{flalign*}
According to Equations \eqref{eq:env_wps_iso} and \eqref{eq:env_bias_iso}, we arrive at
\begin{flalign*}
\epsilon(w,\delta) = 2\left(1-C^W(w,\delta)\right), &&
\end{flalign*}
where $C^W(w,\delta)$ is defined as
\begin{flalign}
\label{eq:env_wcc_iso}
C^W(w,\delta) \equiv \frac{\langle W_\mathrm{gas}(w,\mathbf{r})W_\mathrm{dm}(w,\mathbf{r}) \rangle_{\delta(\mathbf{r})=\delta}}{\sqrt{P^W_\mathrm{gas}(w,\delta)P^W_\mathrm{dm}(w,\delta)}}, &&
\end{flalign}
namely, the env-WCC between the gas and dark matter fields. It is a measure of the statistical coherence between  these two fields, and takes values between $-1$ and $1$. If $C^W(w,\delta)=1$, then the fields are totally correlated, hence the dark matter can be fully determined from the gas by $b^W(w,\delta)$. On the other hand, if $C^W(w,\delta)=-1$, then the fields are totally anticorrelated. Apparently, once $b^W(w,\delta)$ and $C^W(w,\delta)$ are known, the characteristics of the dark matter field can be constrained more accurately from the baryonic observations.

Notice that the above statistics are built on the assumption of an isotropic matter distribution. Generally, the anisotropic statistics as functions of the scale vector $\mathbf{w}$ and the local overdensity $\delta$ can be derived from the anisotropic CW-GDW transform in the same way. For instance, the anisotropic env-WPS, env-bias, and env-WCC are
\begin{flalign}
\label{eq:env_wps_aniso}
P^W_\mathrm{i}(\mathbf{w},\delta)\equiv\Big\langle |W_\mathrm{i}(\mathbf{w},\mathbf{r})|^2 \Big\rangle_{\delta(\mathbf{r})=\delta}, &&
\end{flalign}
\begin{flalign}
\label{eq:env_bias_aniso}
b^W(\mathbf{w},\delta)\equiv\sqrt{\frac{P^W_\mathrm{gas}(\mathbf{w},\delta)}{P^W_\mathrm{dm}(\mathbf{w},\delta)} }, &&
\end{flalign}
\begin{flalign}
\label{eq:env_wcc_aniso}
C^W(\mathbf{w},\delta) \equiv \frac{\langle W_\mathrm{gas}(\mathbf{w},\mathbf{r})W_\mathrm{dm}(\mathbf{w},\mathbf{r}) \rangle_{\delta(\mathbf{r})=\delta}}{\sqrt{P^W_\mathrm{gas}(\mathbf{w},\delta)P^W_\mathrm{dm}(\mathbf{w},\delta)}}, &&
\end{flalign}
respectively. By taking the average of these anisotropic statistics over all $\mathbf{w}$ with the same modulus, they will reduce to the isotropic ones defined by Equations \eqref{eq:env_wps_iso}, \eqref{eq:env_bias_iso}, and \eqref{eq:env_wcc_iso}, i.e.:
\begin{flalign*}
\mathcal{P}^W_\mathrm{i}(w,\delta) & = \Big\langle |W_\mathrm{i}(\mathbf{w},\mathbf{r})|^2 \Big\rangle_{w,\delta} &\sim & \ \ \ P^W_\mathrm{i}(w,\delta) \\
\mathcal{b}^W(w,\delta) & =\sqrt{\frac{\mathcal{P}^W_\mathrm{gas}(w,\delta)}{\mathcal{P}^W_\mathrm{dm}(w,\delta)}} &\sim & \ \ \ b^W(w,\delta) \\
\mathcal{C}^W(w,\delta) & = \frac{\langle W_\mathrm{gas}(\mathbf{w},\mathbf{r})W_\mathrm{dm}(\mathbf{w},\mathbf{r}) \rangle_{w,\delta}}{\sqrt{\mathcal{P}^W_\mathrm{gas}(w,\delta)\mathcal{P}^W_\mathrm{dm}(w,\delta)}} & \sim & \ \ \ C^W(w,\delta).
\end{flalign*}
Since we only care about the dependence of the clustering on the scale modulus $w$ and the local density $\delta$, and also for the sake of minimizing computational effort, we will use those statistics based on the isotropic CW-GDW.\footnote{In Appendix \ref{sec:globalWPSs}, we make comparisons between global WPSs based on the isotropic and anisotropic CW-GDW.}

\section{Results}
\label{sec:results}
We present our main results for the dark matter and baryonic gas clustering at redshifts $0\leqslant z \leqslant 3$. In particular, we explore how the dark matter and gas clustering depends on the environment by computing the env-WPS. We also quantify the deviation between the dark matter and gas distributions by computing the env-bias and env-WCC. In all the results below, the wavelet scale $w$ is expressed as the Fourier wavenumber $k$ according to the relation given by Equation \eqref{eq:scale_wavenumber_iso}. To ensure that our results are unaffected by aliasing effects, we only consider the scales of $w<c_wk_\mathrm{Nyq}/2$, where $k_\mathrm{Nyq}=1024\pi/L_\mathrm{box}$ is the Nyquist frequency.

\subsection{The env-WPS of the density field}

\begin{figure}[t]
	\centerline{\includegraphics[width=0.49\textwidth]{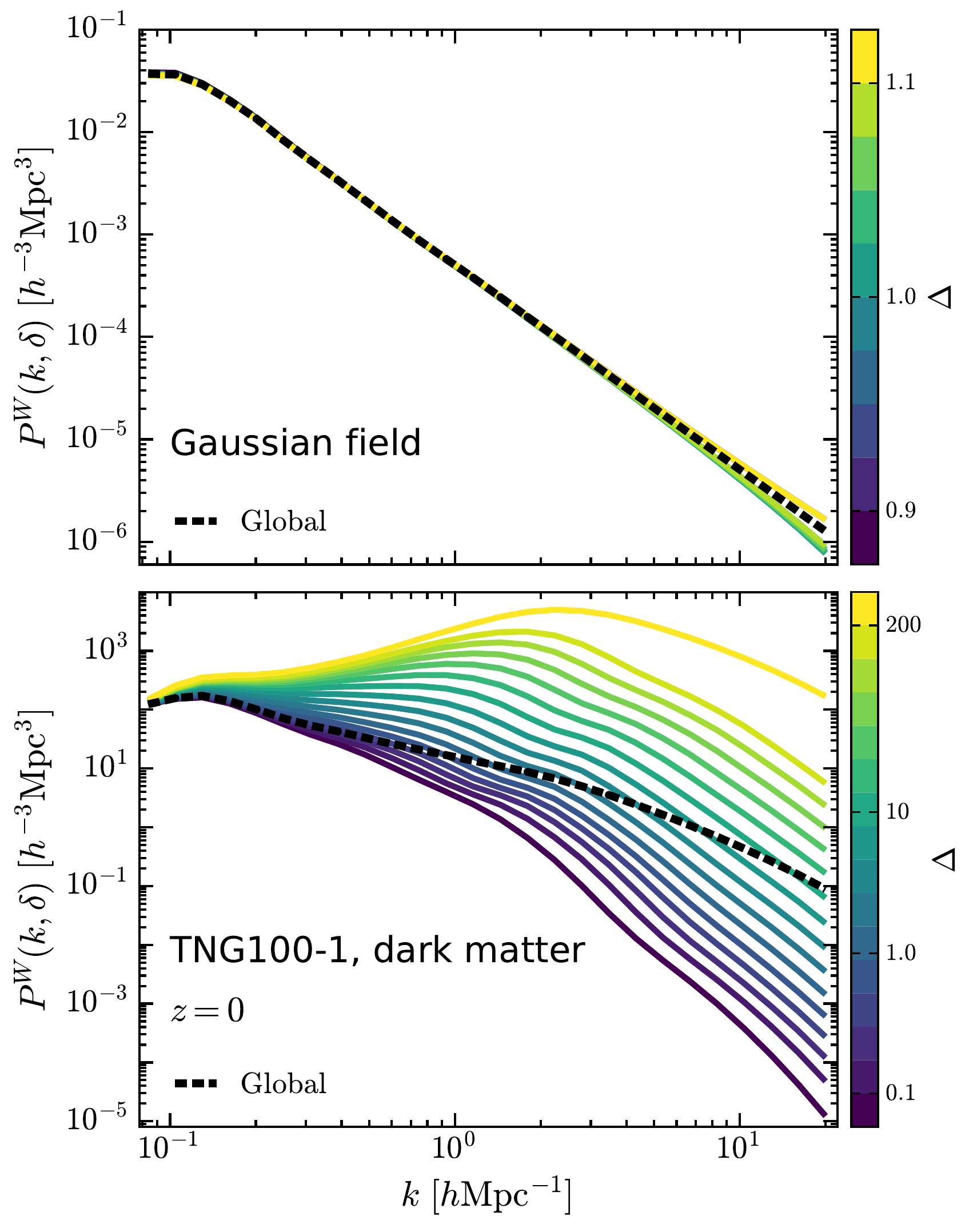}}
	\caption{Comparison of the env-WPS of the Gaussian density field and that of the non-Gaussian density field. \textit{Top panel}: the env-WPS of the Gaussian density field with power-law power spectrum $P(k)\propto k^{-2}$. \textit{Bottom panel}: the env-WPS of the \texttt{TNG100-1} dark matter density field at $z=0$, which is highly non-Gaussian. The environment is defined according to the density division scheme described in Section \ref{sec:measure_procedure}, in descending order of the density from top to bottom.The global WPS is plotted as a dashed line in each panel for reference.}
	\label{fig:env_wps_gaussian_field}
\end{figure}

\begin{figure*}
	\centerline{\includegraphics[width=0.98\textwidth]{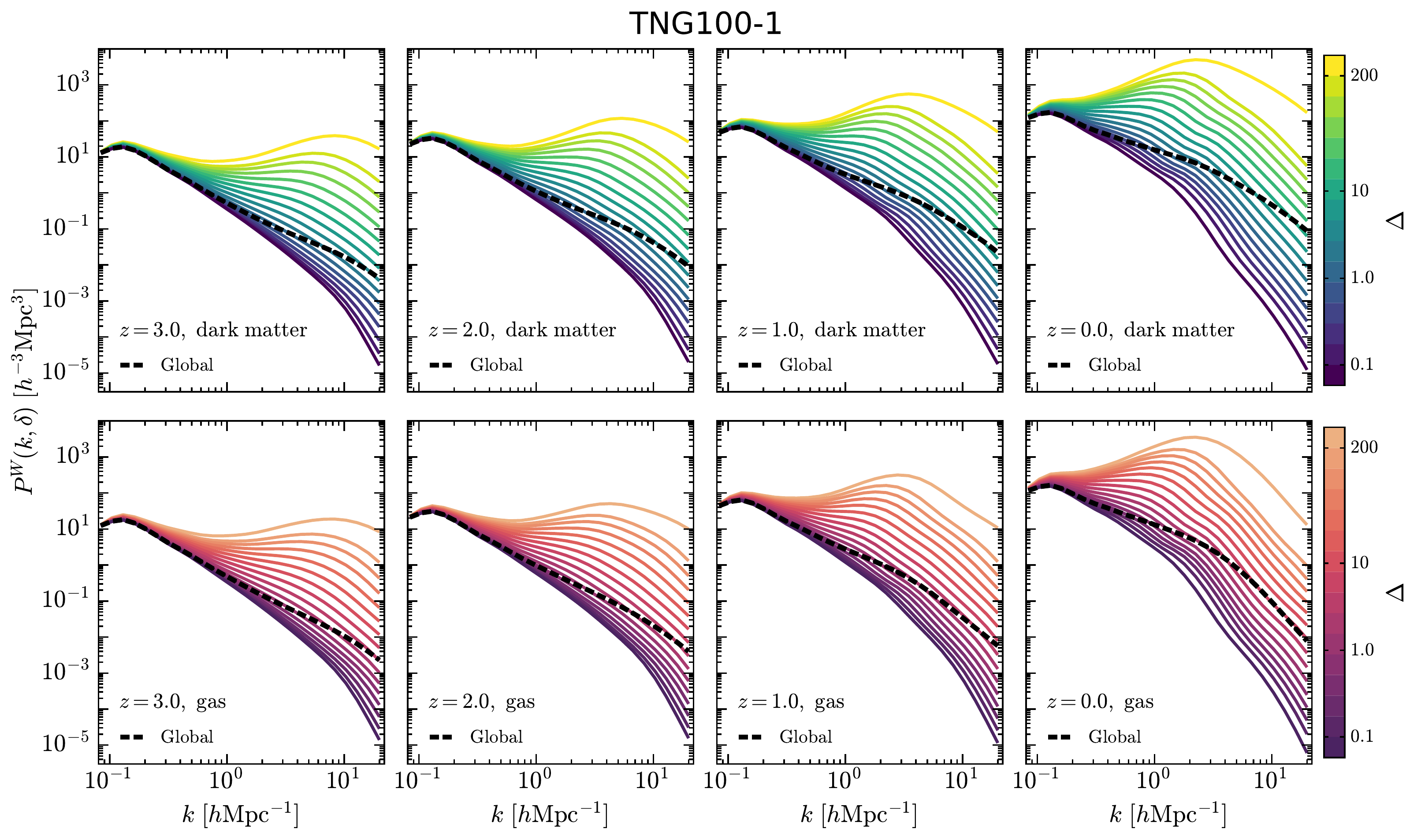}}
	\caption{The env-WPSs of the dark matter (top) and baryonic gas (bottom) density fields of \texttt{TNG100-1} at $z=3$, $2$, $1$ and $0$. The solid lines show the results for all the environments as in Fig.~\ref{fig:env_wps_gaussian_field}. In each panel, the global WPS (dashed black line) is plotted for comparison.}
	\label{fig:evolution_envWPS_TNG100}
\end{figure*}

In Fig.~\ref{fig:env_wps_gaussian_field}, we compare the env-WPSs of a statistically homogeneous Gaussian random field with power-law spectrum $P(k)\propto k^{-2}$ and the \texttt{TNG100-1} dark matter density field at redshift $z=0$. For the Gaussian field, which is completely characterized by the Fourier power spectrum, we can see that its env-WPS shows no environmental dependence. However, for the dark matter field at $z=0$, the env-WPS exhibits a strong dependence on the density environment, confirming that this field is highly non-Gaussian. We show the evolution of the environmental dependence of matter clustering with redshift in Fig.~\ref{fig:evolution_envWPS_TNG100}, in which we focus on the env-WPSs of the dark matter and gas at $z=3.0$, $2.0$, $1.0$, and $0.0$. Now let us consider the behavior of the dark matter. At $z=3$, WPSs in different environments already exhibit different clustering strengths, except that on the largest scales, where they all converge to the global WPS. Specifically, the env-WPS is monotonically increases toward higher-density environments on almost all scales. This result is intuitive, because high-density environments contain more gravitationally bound objects (e.g. halos and subhalos) than low-density environments \citep[e.g.][]{Maulbetsch2007}. At lower redshifts, the env-WPS continues to maintain the same dependence on the density environment with enhanced amplitude, mainly due to nonlinear gravitational effects. For the gas component, its env-WPS follows the behaviors of the dark matter. However, on intermediate and small scales, the env-WPSs of the gas are suppressed, due to baryonic processes resisting the collapse of gas, particularly due to AGN feedback, which heats and expels gas from halos to large radii \citep[e.g.][]{vanDaalen2019}. 

The differences between the dark matter and gas fields can be precisely investigated by the env-bias and env-WCC, which are exhibited in the following.

\begin{figure*}[t]
	\centerline{\includegraphics[width=0.99\textwidth]{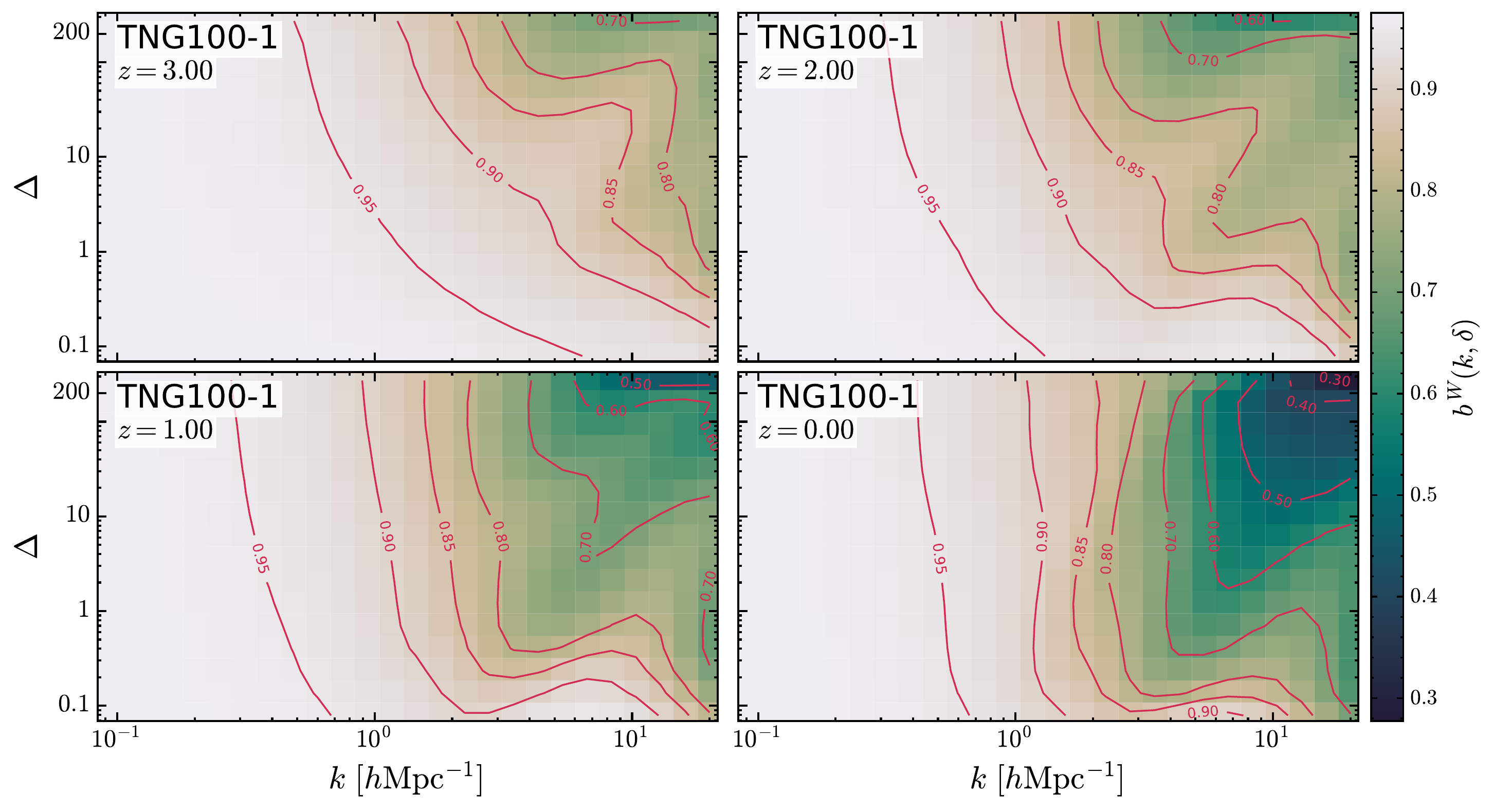}}
	\caption{The env-bias of gas with respect to dark matter defined by equation \eqref{eq:env_bias_iso} at redshifts $z=3$, $2$, $1$ and $0$ measured from the \texttt{TNG100-1} simulation. The bias is roughly constant on large scales with $b^W\sim 0.977$ at $z=3$, $b^W\sim 0.967$ at $z=2$, $b^W\sim 0.964$ at $z=1$ and $b^W\sim 0.973$ at $z=0$.}
	\label{fig:evolution_envbias_TNG100}
\end{figure*}
\begin{figure*}
	\centerline{\includegraphics[width=0.99\textwidth]{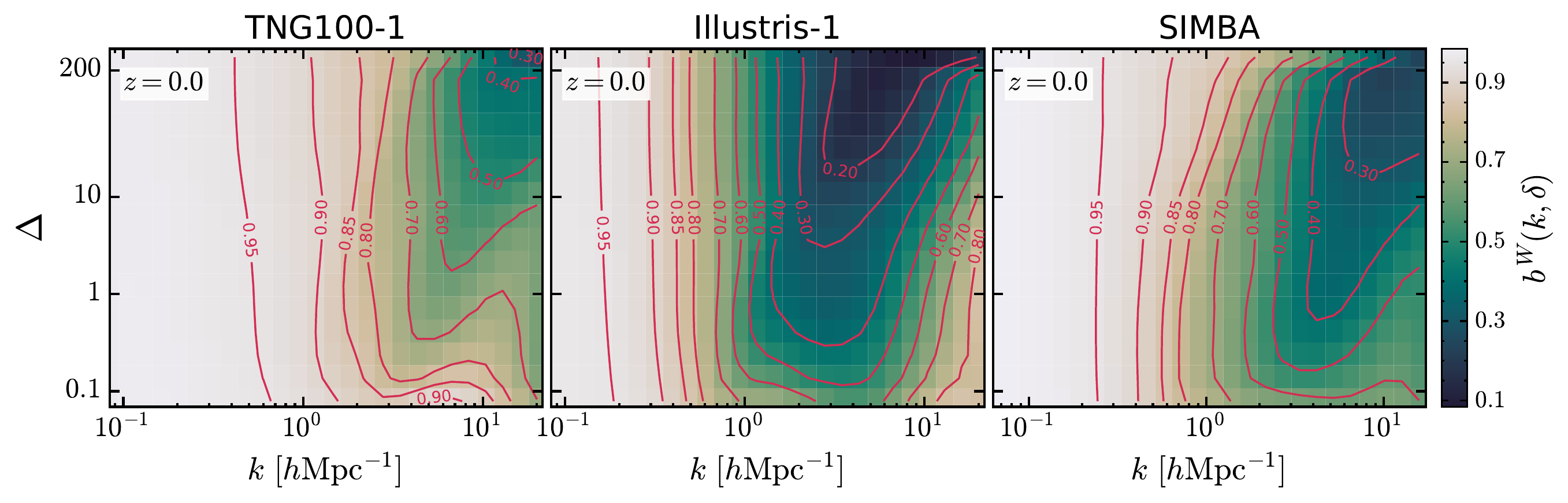}}
	\caption{The env-bias of gas at $z=0$ measured from three simulations: \texttt{TNG100-1}, \texttt{Illustris-1} and \texttt{SIMBA}.}
	\label{fig:envbias_multi_simulations}
\end{figure*}
\begin{figure*}
	\centerline{\includegraphics[width=0.99\textwidth]{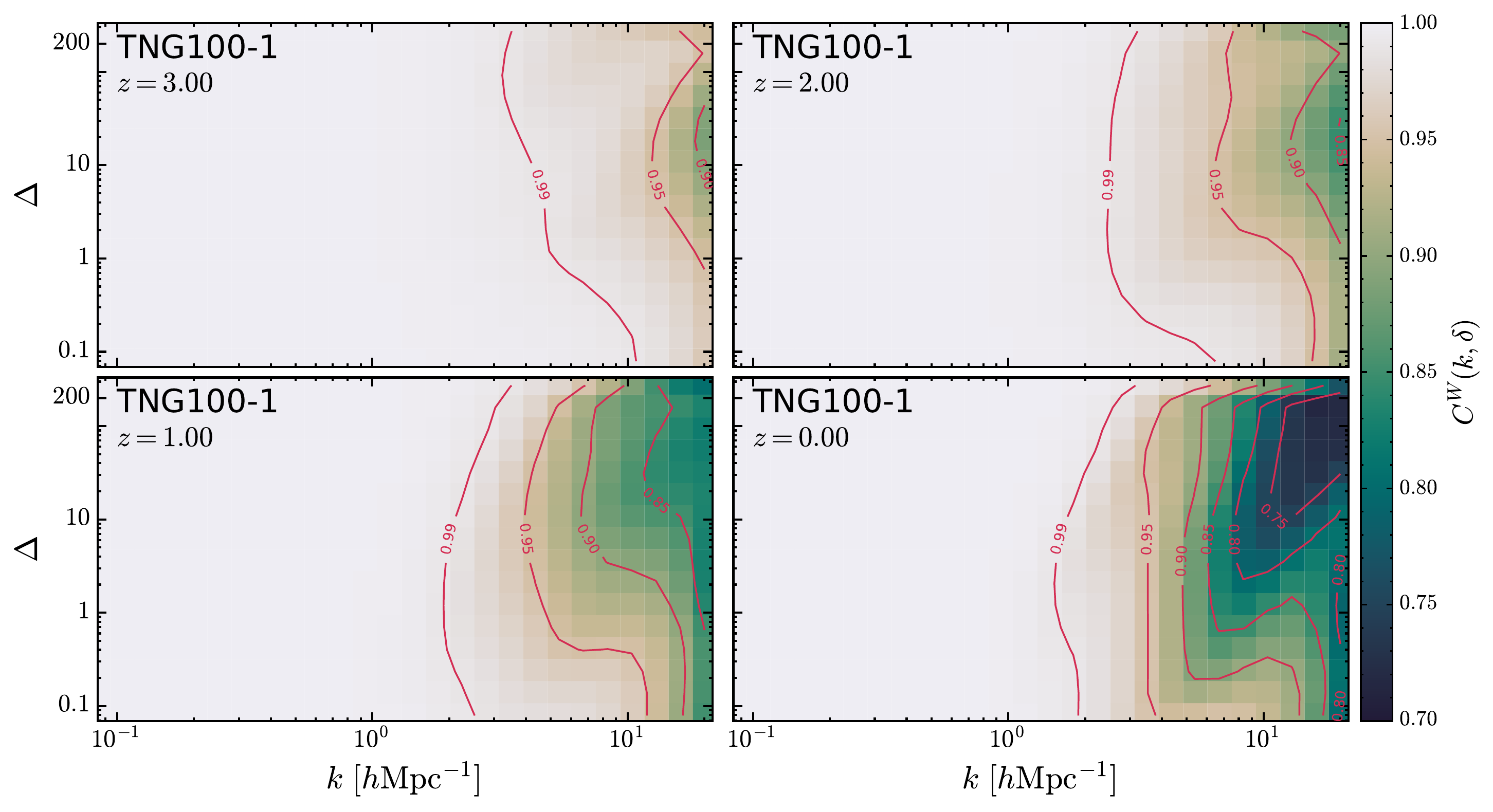}}
	\caption{The env-WCC between gas and dark matter defined by equation \eqref{eq:env_wcc_iso} at redshifts $z=3$, $2$, $1$ and $0$ measured from the \texttt{TNG100-1} simulation. The env-WCC approaches unity on large scales, indicating that the gas and dark matter are coherent with each other.}
	\label{fig:evolution_envwcc_TNG100}
\end{figure*}
\begin{figure*}
	\centerline{\includegraphics[width=0.99\textwidth]{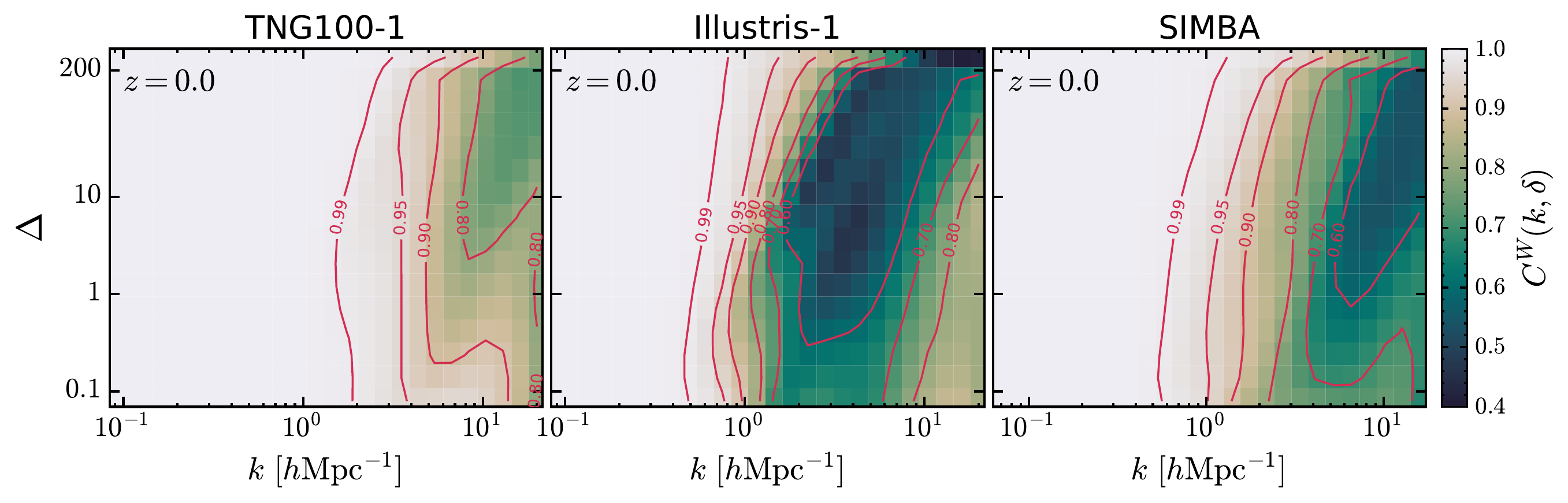}}
	\caption{The env-WCC between gas and dark matter at $z=0$ measured from three simulations: \texttt{TNG100-1}, \texttt{Illustris-1} and \texttt{SIMBA}.}
	\label{fig:envwcc_multi_simulations}
\end{figure*}

\subsection{The Env-bias of the Gas Relative to the Dark Matter}

We now turn to consider the env-bias of the gas relative to the dark matter, the results of which are shown in Fig.~\ref{fig:evolution_envbias_TNG100}. We observe that the environmental bias function $b^W(k,\delta)$ is almost independent of the scale and density environment on very large scales at all epochs. As the scale gets smaller, the environmental dependence of the bias becomes increasingly prominent. The main visible feature is the gas being most severely biased in the densest environment ($\Delta > 200$), and at small scales for all redshifts, while the gas in the extreme underdense environment ($\Delta < 0.1$) shows only minor bias. In more detail, we see that at $z=3$, the env-bias value decreases monotonically as the density increases on scales $0.5\lesssim k\lesssim 10\ h\mathrm{Mpc}^{-1}$, which is consistent with higher gas pressure in denser environments. Note that at this redshift, the suppression of the gas clustering is largely due to the gas pressure, while feedbacks play a minor role \citep[see][]{vanDaalen2011, Chisari2018, Foreman2020}. This monotonically decreasing trend of the env-bias with increasing density becomes weaker toward lower redshifts, and eventually, at $z=0$, the env-bias hardly varies with the density on scales $k\lesssim1\ h\mathrm{Mpc}^{-1}$, and even shows an upturn in densities of $\Delta \gtrsim 10$ around scale of $k\sim 3\ h\mathrm{Mpc}^{-1}$. We also notice that in these densities, the contours of $b^W\gtrsim0.8$ are shifted to smaller scales, compared to the case of $z=1$. Since, at $z<3$, the AGN feedback causes most of the suppression on scales of $k\lesssim10\ h\mathrm{Mpc}^{-1}$, the redshift evolution of the env-bias in this scale range may reflect the AGN feedback efficiency in denser environments being reduced more severely than in less dense environments at lower redshifts. Hence, the gas in denser environments can be preferentially reaccreted to massive halos, confirming the findings of previous research. For example, by measuring the ratio between the total-matter power spectra in the \texttt{Horizon-AGN} and \texttt{Horizon-noAGN} simulations, \citet{Chisari2018} found that the clustering suppression at scale of $k\sim 2\ h\mathrm{Mpc}^{-1}$ diminished slightly from $z=1$ to $0$, and ascribed it to the AGN feedback being less efficient in the most massive halos at lower redshifts. \citet{Foreman2020} measured the ratio of the bispectra between the hydro and dark matter-only simulation runs, and found that it was enhanced at $z=0$, $k\sim 3\ h\mathrm{Mpc}^{-1}$ in \texttt{TNG300-1}, \texttt{TNG100-1} and \texttt{EAGLE}, which was also due to the less effective AGN feedback.
	
On the galactic scales, i.e. $k\gtrsim 10\ h\mathrm{Mpc}^{-1}$, the gas clustering is affected by the joint actions of gas cooling, star formation, and feedbacks, the former two of which are subdominant at the scales we are concerned with ($k\lesssim 20\ h\mathrm{Mpc}^{-1}$). As expected, the env-bias is approximately enhanced with increasing overdensity, indicating that denser environments experience more violent baryonic processes at the redshifts we consider. This result is consistent with the fact that denser environments host higher fractions of quenched galaxies \citep[see, e.g.,][]{Hoyle2012,Moorman2016}.

To determine whether our results depend on the particular galaxy formation model included in the \texttt{TNG100-1}, we also measure the env-bias at $z=0$ from the \texttt{Illustris-1} and \texttt{SIMBA} simulations, which have similar box sizes to the \texttt{TNG100-1}. In Fig.~\ref{fig:envbias_multi_simulations}, the upturn in densities of $\Delta \gtrsim 10$ around the scale of $3 \ h\mathrm{Mpc}^{-1}$ is also observed in \texttt{SIMBA}, which is even more obvious than that in \texttt{TNG100-1}. \texttt{Illustris-1} does not show this feature, which may be due to its overefficient AGN feedback for massive halos \citep{Genel2014}. For the same reason, the gas in \texttt{Illustris-1} suffers the most severe suppression (particularly $b^W\sim 0.2$ in $\Delta \gtrsim 10$ and at $k\gtrsim 2\ h\mathrm{Mpc}^{-1}$) compared to the other simulations. Nonetheless, all simulations agree with the gas being more strongly biased in denser environments and on smaller scales at $z=0$.

\subsection{The Env-WCC between the Dark Matter and the Gas}

The time evolution of the env-WCC vs. scale and density is shown in Fig.~\ref{fig:evolution_envwcc_TNG100}. Overall, we see that the env-WCC is very close to $1$ on large scales for all redshifts, which means that the distributions of the gas and the dark matter are highly coherent on large scales, and therefore the env-bias alone would be sufficient to derive the dark matter field from the gas field.  As the scale decreases, the deviation of the env-WCC from unity becomes larger and shows a stronger environmental dependence, which is due to the redistribution of gas caused by baryonic physical processes.

As revealed by the $C^W=0.99$ contour, the gas and the dark matter in the densest environment ($\Delta \gtrsim 200$) always keep a very high correlation at scales $k\lesssim 3 \ h\mathrm{Mpc}^{-1}$ for all redshifts. We notice that \citet{Farahi2022} reached a similar conclusion: in the \texttt{TNG100-1} simulation, the correlation between the dark matter and gas density profiles of halos is very close to $1$ on scales of $r>R_{200}$, where $R_{200}$ is the radius of a sphere whose enclosed average overdensity is 200 times the critical density. However, the $0.99$ contour in less dense environments is shifted to larger scales at later times.

It can also be seen at $z=3$ that the env-WCC shows an obvious nonmonotonic dependence on overdensity at scales of $k\gtrsim 10 \ h\mathrm{Mpc}^{-1}$, and is the lowest ($C^W<0.90$) between $\Delta \sim 10$ and $\Delta \sim 40$. This least-correlated region in the $k-\delta$ plot gradually expands to larger scales and lower-density environments with decreasing redshift, but is still concentrated in the environments of $0.1 \lesssim \Delta \lesssim 200$ (roughly the filaments and sheets). The gas always correlates tightly with the dark matter in the extreme underdense environments ($\Delta \sim 0.1$) from $z=3$ to $0$. \citet{Yang2021} also reported this feature and pointed out that the gas and the dark matter correlate more tightly in voids.

In Fig.~\ref{fig:envwcc_multi_simulations}, we compare the env-WCC measured from the \texttt{TNG100-1} with those measured from the \texttt{Illustris-1} and \texttt{SIMBA} at $z=0$. All of them show similar trends, with some discrepancies, possibly attributed to different galaxy formation models. Specifically, the gas correlates less with the dark matter in \texttt{Illustris-1}, and in the most underdense environments ($\Delta <0.1 $), the env-WCC is even less than $0.80$ at scales of $1\lesssim k\lesssim 10 \ h\mathrm{Mpc}^{-1}$, whereas this correlation between the gas and the dark matter is largely enhanced in the \texttt{TNG100-1}, with an updated AGN feedback model. The characteristics of the env-WCC in \texttt{SIMBA} are intermediate between those in \texttt{Illustris-1} and \texttt{TNG100-1}.

\section{Discussion and conclusions}

In this paper, we propose three 3D wavelet-based statistical tools---(1) the env-WPS; (2) env-bias; and (3) env-WCC---which are defined as Equations \eqref{eq:env_wps_iso}, \eqref{eq:env_bias_iso} and \eqref{eq:env_wcc_iso}, respectively. These statistics are constructed from the continuous wavelet coefficients, which simultaneously retain spatial and scale information. Therefore, we expect them to be able to quantify the density environment effects on the matter clustering at various scales. To verify this, we apply them to the dark matter and gas density fields of three state-of-the-art cosmological simulations: \texttt{TNG100-1}, \texttt{Illustris-1} and \texttt{SIMBA}. To achieve a better spectral analysis, we use the CW-GDW as our analytic wavelet. Compared to the GDW used in our previous works \citep{Wang2021a, Wang2021b}, this new wavelet has a higher resolution in the frequency domain, and therefore the global WPS based on it is much closer to the Fourier power spectrum, particularly on small scales (see Appendix \ref{sec:globalWPSs}).

This study has confirmed that the env-WPS, env-bias, and env-WCC are able to correctly characterize the dependence of the matter clustering on both the scale and the local density environment. The comparison of the env-WPSs between a Gaussian field and the present-day dark matter field shows that the env-WPS is fully capable of detecting non-Gaussianity (see Fig.~\ref{fig:env_wps_gaussian_field}). Thus, it is promising that the env-WPS will be able to constrain cosmological parameters more tightly. The env-WPSs of the dark matter and gas exhibit similar features: they both converge to the global WPS at large scales, while at small scales they increase with increasing density at $z=3$, $2$, $1$, and $0$ (see Fig.~\ref{fig:evolution_envWPS_TNG100}). 

The differences of the spatial distributions between the gas and the dark matter are quantified by the env-bias and env-WCC (Figs.~\ref{fig:evolution_envbias_TNG100}-\ref{fig:envwcc_multi_simulations}). We notice that at redshift $z=3$, the env-bias decreases with increasing density up to scales of a few times $0.1 \ h\mathrm{Mpc}^{-1}$ in \texttt{TNG100-1}, indicating that the gas clustering is more suppressed in denser environments. At later times, when the AGN feedback effect is significant, this decreasing trend becomes progressively weaker, and eventually an increasing trend emerges at $z=0$, i.e. the env-bias shows an upturn around $k\sim 3 \ h\mathrm{Mpc}^{-1}$ and in $\Delta \gtrsim 10$, which is also observed in \texttt{SIMBA} but not in \texttt{Illustris-1}. This feature is likely a consequence of the decreased AGN feedback strength occurring in high-density environments at later epochs, which is supported by other research \citep{Chisari2018, Foreman2020}.

Moreover, by measuring the env-WCC in \texttt{TNG100-1}, we find that the dark matter and the gas in both the extreme dense ($\Delta \gtrsim 200$) and underdense ($\Delta \lesssim 0.1$) environments always maintain very high correlation at all redshifts over a wide scale range, which qualitatively agrees with the findings in \citet{Farahi2022} and \citet{Yang2021}. Particularly at the present epoch, the env-WCC is greater than $0.9$ in densities of $\Delta \gtrsim 200$ and $\Delta \lesssim 0.1$ at scales of $k \lesssim 10 \ h\mathrm{Mpc}^{-1}$. However, the dark matter and the gas correlate less well in density environments of $0.1 < \Delta < 200$. A similar dependence is also observed in \texttt{Illustris-1} and \texttt{SIMBA}, but with lower env-WCC values, possibly attributed to different galaxy formation models.

In general, our results suggest that density environment has a non-negligible impact on the characteristics of the matter clustering at low redshifts. Even on large scales of $k \lesssim 1 \ h\mathrm{Mpc}^{-1}$, there is a visible density dependence for the env-bias. The env-WCC also shows the density dependence up to scales of $2 \ h\mathrm{Mpc}^{-1}$, and this is even larger in \texttt{Illustris-1} and \texttt{SIMBA}. Consequently, by means of the env-bias and the env-WCC, the dark matter distribution should be predicted more precisely from the baryonic gas, which could be observed by upcoming sky surveys, e.g. CHIME \citep{Bandura2014}, HIRAX \citep{Newburgh2016} and SKA \citep{Bacon2020}. In this sense, more cosmological simulations are needed to conduct a detailed census of the env-bias and the env-WCC for dark matter and gas. Furthermore, considering different gas phases, such as the warm-hot intergalactic medium or the diffuse intergalactic medium, may be more meaningful. Additionally, it is also important to investigate the differences between other tracers, such as halos and galaxies, and the dark matter density field from our wavelet statistics. These issues are left for future investigations.

\section*{Acknowledgments}

The authors thank the anonymous referee for helpful comments and suggestions. The authors also thank the Illustris, IllustrisTNG, and SIMBA teams for their publicly available data.

P.H. acknowledges the support from the National Science Foundation of China (No. 12047569, 12147217), and from the Natural Science Foundation of Jilin Province, China (No. 20180101228JC). 

\software{NumPy (\url{https://numpy.org/}), pyFFTW (\url{https://github.com/pyFFTW/pyFFTW}), SciPy (\url{https://scipy.org/}), powerbox (\url{https://powerbox.readthedocs.io/en/latest/}), Matplotlib (\url{https://matplotlib.org/}), Mayavi (\url{https://docs.enthought.com/mayavi/mayavi/}), Jupyter Notebook (\url{https://jupyter.org/})}

\appendix
\restartappendixnumbering
\section{Comparison between the global WPS and the conventional Fourier power spectrum in 2D}
\label{sec:globalWPSs}
\begin{figure*}
	\centerline{\includegraphics[width=0.76\textwidth]{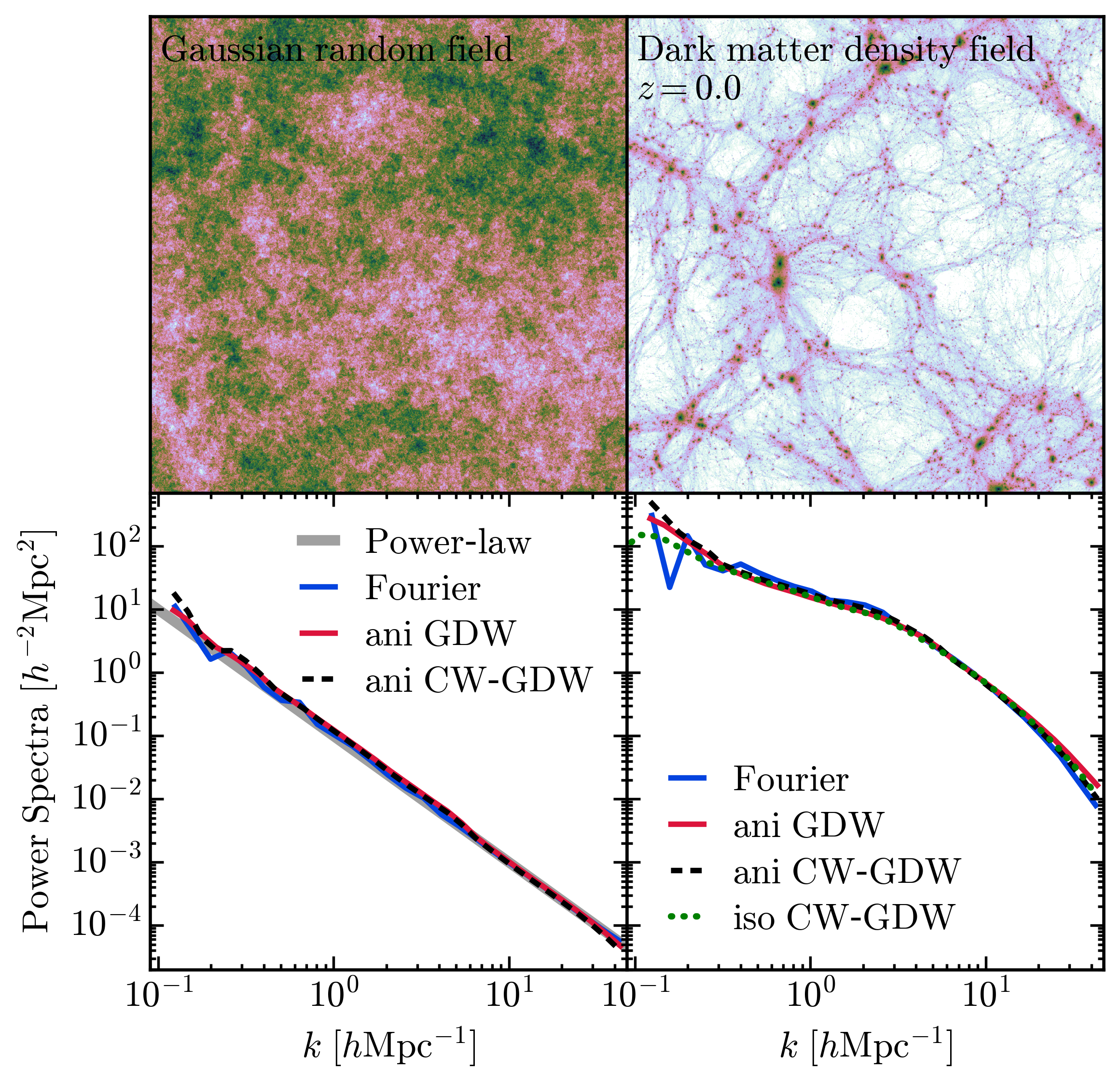}}
	\caption{Comparison of Fourier power spectra and global wavelet power spectra. \textit{Top left}: 2D Gaussian random field with power-law power spectrum $P(k)\propto k^{-2}$. \textit{Bottom left}: the measured power spectra of the random field. \textit{Top right}: 2D projected dark matter density field at redshift $z=0$ of \texttt{TNG100-1} in a slice of width $75 \ h^{-1}\mathrm{Mpc}$ and thickness $16\ h^{-1}\mathrm{Mpc}$. \textit{Bottom right}: the measured power spectra of the dark matter density field. In these two bottom panels, all wavelet power spectra are normalized to fit the Fourier power spectrum.}
	\label{fig:global_spectra}
\end{figure*}
Until now, we have designed two kinds of wavelets, i.e. the GDW and the CW-GDW. Here, we check whether these wavelets could reproduce the Fourier power spectrum, which is important to do, since most works on matter clustering are done by Fourier analysis. To do this, we create a 2D Gaussian random field with a given power-law power spectrum $P(k)=0.1k^{-2}$ and make a comparison between its measured Fourier and global WPS, based on the anisotropic GDW and CW-GDW, which are computed by averaging all the wavelet coefficients over the whole space and the scale vector $\mathbf{w}$ with the same modulus. As shown in the bottom left panel of Fig. \ref{fig:global_spectra}, both the WPS with anisotropic GDW and CW-GDW, as well as Fourier power spectrum, converge to the power-law power spectrum. Then we measure the Fourier and global wavelet power spectra of the 2D density field of the dark matter at $z=0$, the results for which are shown in the bottom right panel of Fig.~\ref{fig:global_spectra}. It can be seen that all the power spectra have qualitatively the same shape and amplitude, except at the large scales , which suffer from the finite-volume
effect. Therefore, the WPS can also correctly characterize the matter distribution. 

Nevertheless, the use of different analytic wavelets can lead to subtle differences in the results, which can be explained by the fact that the global WPS is the output of the Fourier power spectrum being smoothed by the square of a wavelet kernel \citep[see, e.g.,][]{Wang2021b}. Obviously, the power-law spectrum is immune to such a smoothing operation, but, in general, the shape of the WPS will be smoother than the corresponding Fourier power spectrum. In our case, the GDW power spectrum is slightly lower than both the Fourier and CW-GDW power spectra on the scales $k\lesssim 3 \ h\mathrm{Mpc}^{-1}$, and higher on $k\gtrsim 20 \ h\mathrm{Mpc}^{-1}$ for the highly nonlinear dark matter density field. This is due to the poor localization of the GDW in the frequency domain, as shown in Fig.~\ref{fig:CW-GDW_1d}. Since the CW-GDW is more concentrated in frequency than the GDW, we see that the WPS based on the CW-GDW is closer to the Fourier power spectrum than that based on the GDW, especially on small scales. Finally, we compare the global WPSs, based on the anisotropic CW-GDW and the isotropic CW-GDW. As expected, they are almost identical.

\bibliography{paper}{}
\bibliographystyle{aasjournal}

\end{document}